\documentclass[twocolumn]{aastex63}
\usepackage{amsmath}

\accepted{March 2, 2022}
\submitjournal{ApJ}
\shorttitle{X-ray/Radio View of massive quiescent galaxies}
\shortauthors{Ito et al.}

\graphicspath{{./}{figures/}}
\begin{document}

\title{COSMOS2020: Ubiquitous AGN Activity of Massive Quiescent Galaxies at $0<z<5$ Revealed by X-ray and Radio Stacking}

\correspondingauthor{Kei Ito}
\email{kei.ito@grad.nao.ac.jp}
\author[0000-0002-9453-0381]{Kei Ito}
\affiliation{Department of Astronomical Science, The Graduate University for Advanced Studies, SOKENDAI, 2-21-1 Osawa, Mitaka, Tokyo, 181-8588, Japan} 
\affiliation{National Astronomical Observatory of Japan, 2-21-1 Osawa, Mitaka, Tokyo, 181-8588, Japan}
\author{Masayuki Tanaka}
\affiliation{National Astronomical Observatory of Japan, 2-21-1 Osawa, Mitaka, Tokyo, 181-8588, Japan}
\affiliation{Department of Astronomical Science, The Graduate University for Advanced Studies, SOKENDAI, 2-21-1 Osawa, Mitaka, Tokyo, 181-8588, Japan}
\author[0000-0002-7562-485X]{Takamitsu Miyaji}
\affiliation{Universidad Nacional Aut\'onoma de M\'exico (UNAM), Instituto de Astronom\'{i}a, AP 106,  Ensenada 22800, BC, M\'exico}
\author{Olivier Ilbert}
\affiliation{Aix Marseille Univ, CNRS, CNES, LAM, Marseille, France}
\author{Olivier B. Kauffmann}
\affiliation{Aix Marseille Univ, CNRS, CNES, LAM, Marseille, France}
\author[0000-0002-6610-2048]{Anton M. Koekemoer}
\affiliation{Space Telescope Science Institute, 3700 San Martin Dr., 
Baltimore, MD 21218, USA}
\author{Stefano Marchesi}
\affiliation{INAF - Osservatorio di Astrofisica e Scienza dello Spazio di Bologna, Via Piero Gobetti, 93/3, 40129 Bologna, Italy}
\affiliation{Department of Physics and Astronomy, Clemson University, Kinard Lab of Physics, Clemson, SC 29634, USA}
\author{Marko Shuntov}
\affiliation{Institut d'Astrophysique de Paris, UMR 7095, CNRS, and Sorbonne Universit\'e, 98 bis boulevard Arago, 75014 Paris, France}
\author{Sune Toft}
\affiliation{Cosmic Dawn Center (DAWN), Denmark}
\affiliation{Niels Bohr Institute, University of Copenhagen, Jagtvej 128, 2200 Copenhagen, Denmark}
\author{Francesco Valentino}
\affiliation{Cosmic Dawn Center (DAWN), Denmark}
\affiliation{Niels Bohr Institute, University of Copenhagen, Jagtvej 128, 2200 Copenhagen, Denmark}
\author{John R. Weaver}
\affiliation{Cosmic Dawn Center (DAWN), Denmark}
\affiliation{Niels Bohr Institute, University of Copenhagen, Jagtvej 128, 2200 Copenhagen, Denmark}

\begin{abstract}
We characterize the average X-ray and radio properties of quiescent galaxies (QGs) with $\log{(M_\star/M_\odot)}>10$ at $0<z<5$. QGs are photometrically selected from the latest COSMOS2020 catalog. We conduct the stacking analysis of X-ray images of the Chandra COSMOS Legacy Survey for individually undetected QGs. Thanks to the large sample and deep images, the stacked X-ray signal is significantly detected up to $z\sim5$. The average X-ray luminosity can not be explained by the X-ray luminosity of X-ray binaries, suggesting that the low-luminosity active galactic nuclei (AGNs) ubiquitously exist in QGs. Moreover, the X-ray AGN luminosity of QGs at $z>1.5$ is higher than that of star-forming galaxies (SFGs), derived in the same manner as QGs. The stacking analysis of the VLA-COSMOS images is conducted for the identical sample, and the radio signal for QGs is also detected up to $z\sim5$. We find that the radio AGN luminosity of QGs at $z>1.5$ is also higher than SFGs, which is in good agreement with the X-ray analysis. The enhanced AGN activity in QGs suggested by the individual analysis in the X-ray and radio wavelength supports its important role for quenching at high redshift. Their enhanced AGN activity is less obvious at $z<1.5$, which can be interpreted as an increasing role of others at lower redshifts, such as environmental quenching.
\end{abstract}

\keywords{}

\section{Introduction} \label{sec:intro}
\par Massive elliptical galaxies in the local universe are often quiescent. Detailed spectroscopic analyses indicate a simple star formation history of these galaxies; they formed in an intense burst of star formation at high redshifts, followed by passive evolution \citep[e.g.,][]{Thomas2005}. 
\par Recent studies have unveiled the existence of massive and compact galaxies with suppressed star formation rates (SFRs) even at high redshift. Several studies have photometrically identified massive quiescent galaxies (QGs) at $z<6$ \citep[e.g.,][]{Straatman2014,Davidzon2017,Merlin2019,Mawatari2020}. Thanks to the high sensitivity of the latest near-infrared spectrograph, they are now spectroscopically confirmed up to $z\sim4$ \citep{Glazebrook2017,Schreiber2018,Tanaka2019,Forrest2020,Valentino2020,DEugenio2020}. Their star formation history inferred from the SED modeling implies their intense star formation followed by quenching in a short time scale, in agreement with star formation histories of massive elliptical galaxies in the local universe.
\par It is not well understood what physical processes are responsible for the rapid quenching and suppressing subsequent star formation activities, although the cold streams are expected to supply gas at the high redshift \citep[e.g.,][]{Dekel2009}. One of the preferred mechanisms is the feedback from active galactic nuclei (AGNs). The radiation, wind, or radio jet from AGNs can eject gas from galaxies or heat the gas in/around galaxies, thereby suppressing star formation activity, although the detailed mechanism still remains unclear. In the local universe, there are multiple lines of observational evidence for the AGN feedback \citep[see][for review]{Fabian2012}. 
\par Investigating the AGN activity of QGs is essential to explore the relation between quenching and AGNs. In that sense, the stacking of the X-ray and radio images is a powerful tool to reveal the average picture of the AGN activity, reaching low luminosities even at high redshift. So far, \citet{Olsen2013} show that the stacked X-ray luminosity of individually undetected QGs at $z\sim2$ cannot be explained by the X-ray luminosity originated from the star formation, lending support to the presence of AGNs. The radio stacking of QGs at $z\leq2$ also suggests a high radio luminosity for their star formation rate \citep{Man2016,Gobat2017,Gobat2018, Magdis2021}, which is in line with the X-ray picture. On the other hand, X-ray and radio properties of typical QGs are almost unexplored at higher redshifts ($z>2$). Indeed, several studies have discussed the X-ray property focusing on the only limited sample. At $3<z<4$, \citet{Schreiber2018} focus only on X-ray detected sources. \citet{Carraro2020} and \citet{DEugenio2020} include individually detected sources in their statistical analysis at $z<3.5$, but this is potentially a problem. Powerful AGNs, which are detected individually, may not sample the entire AGN population \citep[see the X-ray luminosity function work such as][]{Ueda2014,Miyaji2015} and might make the average trend of the sample skewed owing to their high luminosity. This limitation toward the higher redshift is primarily due to the rapid decrease of the number of QGs toward higher redshift \citep[e.g.,][]{Merlin2019} and the lack of deep X-ray and radio imaging which can detect the faint signal of the stacked high redshift objects. In addition, if we want to discuss AGN activity in the context of quenching, we should compare the properties of QGs to those of star-forming galaxies (SFGs).
\par In this study, we perform a systematic stacking analysis for the X-ray and radio images of QGs at $0<z<5$ which are individually undetected in the X-ray. QGs are selected using the latest photometric catalog of the COSMOS field \citep[COSMOS2020,][]{Weaver2021}. The COSMOS field covers $\sim2\deg^2$, and in addition to the deep multiband optical and infrared data, the deep X-ray and the radio data exist \citep{Schinnerer2007, Civano2016,Marchesi2016, Smolcic2017}, enabling us to unveil the average properties of typical QGs with the largest sample out to the highest redshift ever. By comparing them with SFGs selected from the same catalog, we examine the difference in AGN activity between these two different populations.
\par This paper consists of the following sections. In Section \ref{sec:2}, we introduce the COSMOS2020 catalog and our galaxy sample, followed by X-ray stacking analysis in Section \ref{sec:3}. In Section \ref{sec:4}, we perform the same analysis using the radio data. Section \ref{sec:5} discusses the connection between black hole activity and quenching. Finally, we conclude the paper in Section \ref{sec:6}. We assume the following cosmological parameters: $H_0 = 70\ {\rm km\ s^{-1}\ Mpc^{-1}}$, $\Omega_m = 0.3$, and $\Omega_\Lambda = 0.7$. All magnitude are expressed in the AB system \citep{Oke1983}.
\section{Sample Selection} \label{sec:2}
\subsection{Photometric redshift measurement}
\par We use the latest photometric catalog from Cosmic Evolution Survey \citep[COSMOS,][]{Scoville2007}, called COSMOS2020 \citep{Weaver2021}, covering a $\sim2\deg^2$ field. This catalog consists of multi-band photometry from FUV band of GALEX to IRAC photometry of Spitzer Space telescope, which leads to the wavelength coverage of $0.1-10\mu {\rm m}$. This is an updated version of the previous COSMOS multi-photometry catalog of \citet{Laigle2016} (COSMOS2015) and includes the latest imaging data of this field, such as $U$-band data of CLAUDS survey \citep[][]{Sawicki2019} by CFHT MegaCam, $grizy$-band data of HSC-SSP PDR2 \citep[][]{Aihara2019} by Subaru HSC, $YJHKs$-band data of UltraVISTA DR4 \citep[][]{McCracken2012} by VISTA VIRCAM, and Spitzer/IRAC channel 1,2,3, and 4 data of the Cosmic Dawn Survey \citep[][]{Moneti2021}. In this study, we use photometry of the ``classic catalog". In the classic catalog, the aperture photometry is used for all bands except for the IRAC bands. For the IRAC bands, where the source confusion makes the aperture photometry difficult, we use the photometry measured by the IRACLEAN software \citep{Hsieh2012}. We construct the magnitude-limited sample with $Ks<25$ mag from the catalog. In addition, sources in the vicinity of bright stars are removed from the sample based on the bright star mask of HSC-SSP PDR2 \citep[][]{Coupon2018}, which results in the final area of $\sim1.4\ {\rm deg^2}$.
\par We utilize photometric redshifts estimated from the {\tt MIZUKI} code \citep{Tanaka2015}. This code simultaneously derives the photometric redshifts and physical properties using Bayesian priors. An advantage of this approach is that the uncertainty of the photometric redshift is properly included in the uncertainty estimate for other parameters. {\tt MIZUKI} uses spectral templates from \citet{Bruzual2003}, assuming Chabrier IMF \citep{Chabrier2003}, and Calzetti dust attenuation curve \citep{Calzetti2000}. We use an exponentially declining SFR, i.e., ${\rm SFR} \propto \exp(-t/\tau)$, where $t$ is time since the onset of star formation. The $\tau$ is assumed to be $0.1\ {\rm Gyr}<\tau<11\ {\rm Gyr}$ in addition to $\tau=0,\ \infty$, which is equivalent to the single stellar population model and the constant SFR model. The age grid is between 0.05 Gyr and 14 Gyr. Also, the optical depth in the V band ($\tau_{V}$) varies between 0 and 5. Because the templates mentioned above include only stellar emissions, the nebular emissions are included according to \citet{Inoue2011}. We remove sources with unreliable photo-$z$ by applying a reduced chi-square cut of $\chi^2_\nu<50$ and a photo-$z$ $(z_{\rm upper,95}-z_{\rm lower,95})/(1+z)<0.7$, where the term in the brackets is the $95\%$ confidence range. From these cut, only a $0.3\%$ and $4\%$ of total objects are removed, which are too small to alter the properties of the sample. Finally, this study constructs 322,743 galaxies sample at $0<z<5$ in total.
\par
\begin{figure}
    \centering
    \includegraphics[width=8cm]{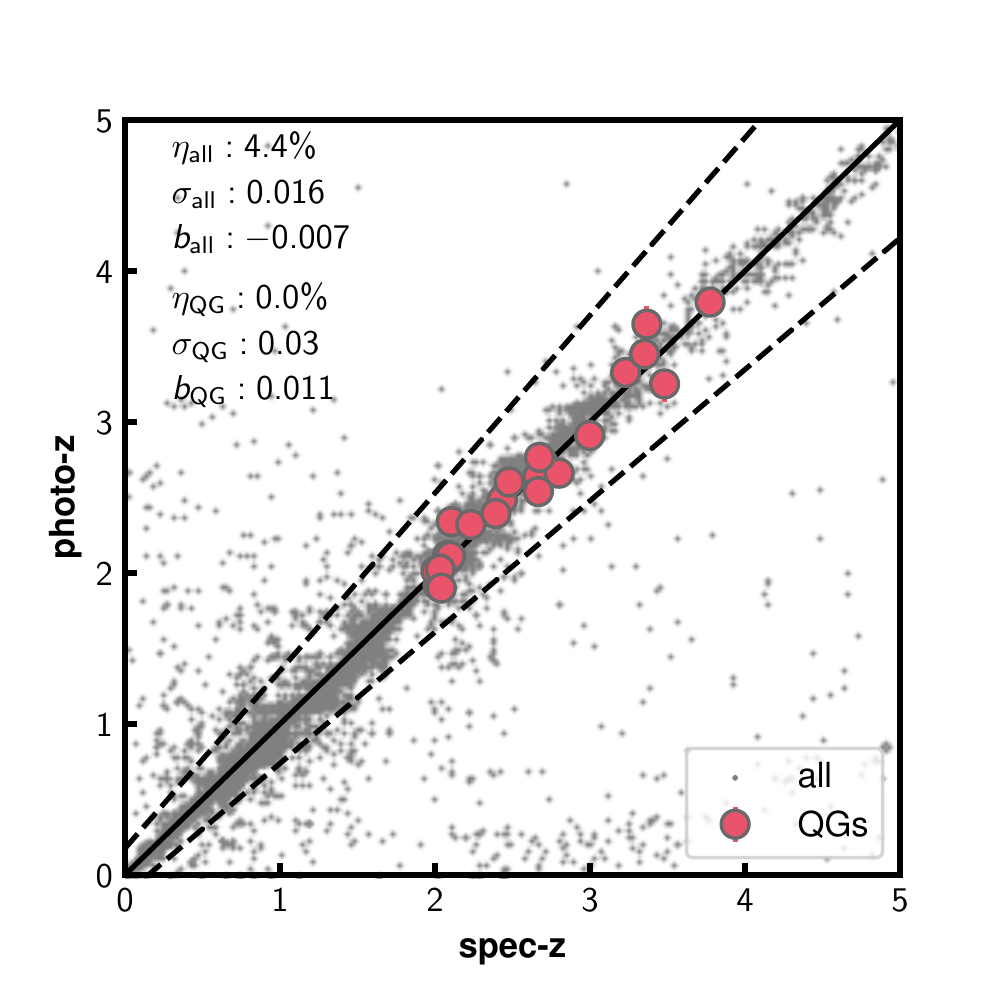}
    \caption{Comparison of the photometric redshift with the spectroscopic redshift of all spectroscopic confirmed objects (gray) and spectroscopic confirmed QGs at $z>2$ \citep[red,][]{Belli2017,Schreiber2018,Saracco2020,Stockmann2020,Valentino2020,DEugenio2020,McConachie2021}. The outlier rate $\eta$, precision measured with the normalized median absolute deviation $\sigma$, and bias $b$ are shown.}
    \label{fig:1}
\end{figure}

 In Figure \ref{fig:1}, our photometric redshifts are compared with all spectroscopic redshifts available in the COSMOS field (Ilbert et al. in prep.). To infer the photometric redshift accuracy, we estimate the outlier rate $\eta$, which is defined as the fraction of objects with $|z_{\rm phot}-z_{\rm spec}|/(1 + z_{\rm spec}) > 0.15$, the precision measured with the normalized median absolute deviation $\sigma_{\rm NMAD}=1.48\times{\rm median}\left(|(z_{\rm phot}-z_{\rm spec})-{\rm median}(z_{\rm phot}-z_{\rm spec})|/(1+z_{\rm spec})\right)$, and the bias $b={\rm median} (z_{\rm phot}-z_{\rm spec})$, where $z_{\rm phot},\ z_{\rm spec}$ are the photometric redshift and spectroscopic redshift, respectively. This comparison shows that our photometric redshift have $\eta=4.4\%$, $\sigma_{\rm NMAD}=0.016$ and $b=-0.007$. In addition, {\tt MIZUKI} also predicts the redshift of high redshift QGs with the good accuracy. The spectroscopically confirmed QG sample is collected from the literature \citep{Belli2017,Schreiber2018,Saracco2020,Stockmann2020,Valentino2020,DEugenio2020,McConachie2021}, which confirm galaxies at $z>2$ with the quiescent features in the near-infrared spectrum (e.g., Balmer absorption lines, Balmer break). The outlier rate, precision, and bias of the redshift for the spec-$z$ QG sample are estimated as $\eta=0.0\%$, $\sigma_{\rm NMAD}=0.03$ and $b=0.011$, respectively.
\subsection{Quiescent Galaxy Selection}
\par There are two popular methods to select QGs. One is based on their colors such as by the $UVJ$ or $NUVrJ$ colors \citep[e.g.,][]{Wuyts2007,Muzzin2013,Straatman2014}. This method is suitable for selecting QGs from the limited information, but it may miss recently quenched galaxies with a short star-formation time scale, which can be more likely exist in the high redshift \citep[e.g.,][]{Merlin2018}. Indeed, some spectroscopically confirmed QGs at $z\sim3-4$ are not satisfied with the criteria of these color selection, even though their star formation rate is indeed suppressed \citep[e.g.,][]{Schreiber2018,Valentino2020}. The other method is based on the specific SFR (sSFR). In this method, we can directly select QGs based on their star formation activities. Here, we select QGs based on the latter method following our previous works \citep{Kubo2018, Tanaka2019, Ito2021}. We select galaxies with $\log{\rm (sSFR_{1\sigma, upper}/{\rm yr^{-1}})}<-11,\ -10.5,\ -10.0,\ -9.5$ as QGs at $0<z<0.5$, $0.5<z<1.0$, $1.0<z<2.0$, $2<z<5$, respectively. Here, ${\rm sSFR_{1\sigma, upper}}$ is the upper limit of sSFR, which is defined as the ratio of the $1\sigma$ upper limit of SFR to the $1\sigma$ lower limit of stellar mass, derived from the SED fitting. This threshold roughly corresponds to $\sim1$ dex below the star formation main sequence in each redshift bin. We classify the other galaxies as SFGs. We note that \citet{Tanaka2015} shows that {\tt MIZUKI} overall provides the stellar mass consistent with that from other photo-$z$ code \citep[{\tt FAST},][]{Kriek2009} and SFR consistent with that from the rest-frame UV and IR luminosity.
\par It is possible that dusty star-forming galaxies contaminate the QG sample. \citet{DEugenio2020} try to remove them using the Spitzer/MIPS $24\mu$m from their QG sample at $z\sim3$. We have confirmed that the main results in this work do not change, even if we remove objects with $S/N>4$ detection in MIPS $24\mu$m \citep{LeFloch2009} (Appendix \ref{sec:B}). On the other hand, we note that $24\mu$m is sensitive not only to star formation but also to AGNs at that redshift. Therefore, we chose not to remove objects detected at $24\mu$m. We also check whether our QGs are detected in 850$\mu$m in the ``super-deblended" catalog by \citet{Jin2018}. The fraction of $S/N>3$ detection is only $0.2\%$ for QG sample, where the median $1\sigma$ flux error of 850$\mu$m is 1.37mJy. Under the assumption of the modified black body dust SED with typical dust temperature of sub-millimeter galaxies \citep{Dudzeviciute2020} and the relation of \citet{Kennicutt1998} corrected to \citet{Chabrier2003} IMF \citep[c.f.,][]{Madau2014}, this $1\sigma$ flux error corresponds to $\sim100M_\odot{\rm yr^{-1}}$ at $z\sim3$. In addition, the radio/X-ray AGN luminosity of QGs follows the local relation of AGNs, after subtracting the luminosity related to the star formation from the observed one (see discussion in Section \ref{sec:5-1} and Figure \ref{fig:15}). These also support that the contamination from dusty star-forming galaxies is not significant.
\subsection{Stacking Subsamples}
\par
\begin{figure*}
    \centering
    \includegraphics[width=16cm]{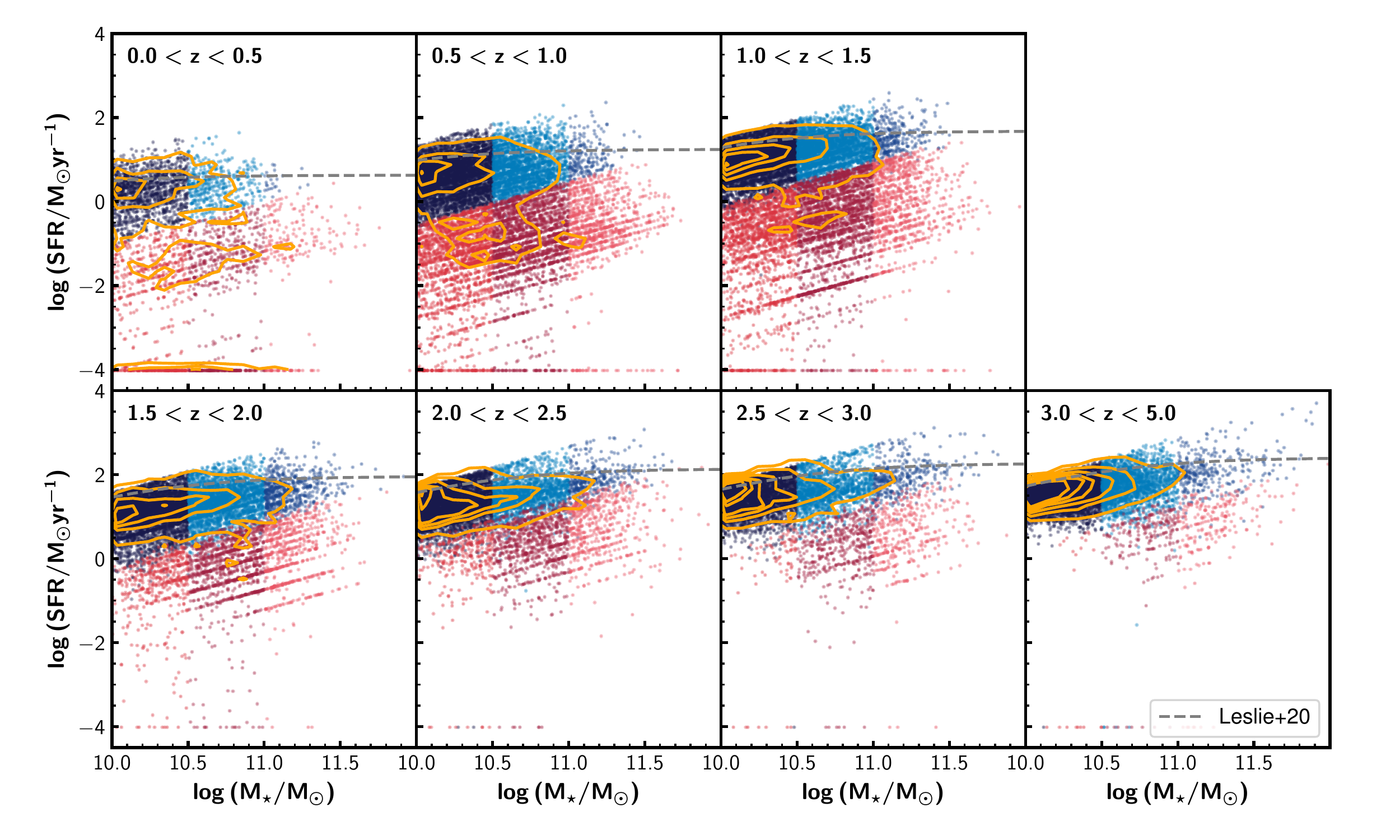}
    \caption{Relation of stellar mass and star formation rate of our sample. The red-colored objects are classified as quiescent galaxies and the others as star-forming galaxies. Different colors in each panel show different stellar mass bins. Objects with SFRs lower than $10^{-4}\ {M_\odot\ {\rm yr^{-1}}}$ are arbitrarily located at SFR of $10^{-4}\ {M_\odot\ {\rm yr^{-1}}}$ only for illustrative purpose. Orange contours show the distribution of all galaxies at that redshift. The slight stripe of data points is due to the model grid of the SED fitting. The gray dashed line shows the best-fit main sequence of star-forming galaxies from \citet{Leslie2020}.}
    \label{fig:2}
\end{figure*}

\begin{figure*}
    \centering
    \includegraphics[width= 16cm]{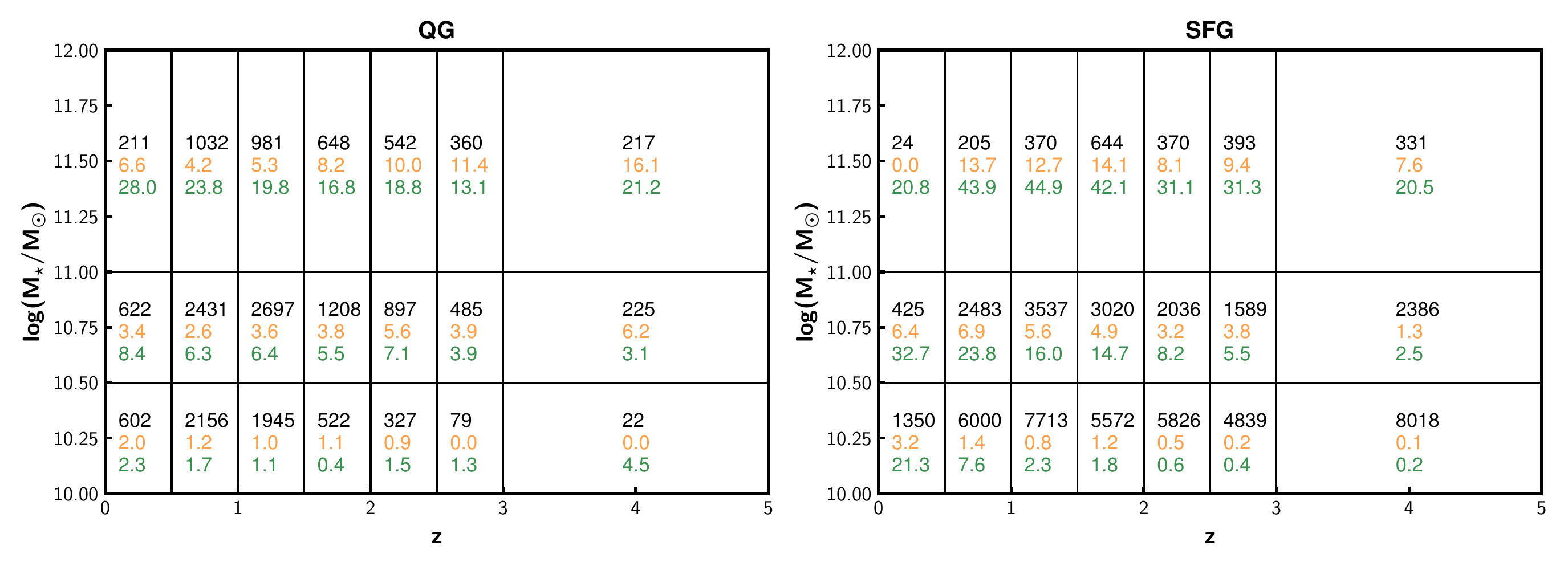}
    \caption{Total number of galaxies (black) in each redshift and stellar mass bin for QGs (left) and SFGs (right). Orange and green numbers show the fraction of sources detected in the X-ray and radio in the percent notation, respectively.}
    \label{fig:3}
\end{figure*}

 We define subsamples of QGs and SFGs based on their stellar mass and redshift. We divide galaxies into three stellar mass bins at $\log{(M_\star/M_\odot)}\geq10.0$ and seven redshift bins at $0<z<5$, as shown in Figure \ref{fig:2}. In order to increase the signal-to-noise ratio in the stacking, the bin size is larger at the most massive bin and the highest redshift bin. Our star formation main sequence is slightly lower than that of \citet{Leslie2020}. On the other hand, the main sequence of \citet{Leslie2020} is also higher than those of the literature \citep[e.g.,][]{Speagle2014,Schreiber2015,Tomczak2016,Iyer2018}. These difference might be due to the sample selection or the SFR estimation method, since we derive the SFR through the SED fitting, whereas \citet{Leslie2020} derive it from the stacked radio luminosity. We summarize the number of sources in each subsample in Figure \ref{fig:3}.
\par The stellar mass completeness due to the magnitude limit cut is calculated from the method in the literature \citep[e.g.,][]{Pozzetti2010,Laigle2016,Davidzon2017}. The magnitude cut of $Ks<25$ mag corresponds to the 90\% completeness limit of QGs with $\log{(M_\star/M_\odot)}=10.2,\ 10.4$ at $2.5<z<3.0$ and $3.0<z<5.0$, respectively. The least massive subsample at these redshifts are thus complete less than 90\%, but the derived flux does not change even if we remove the magnitude cut. We note that other bins are more than $90\%$ complete for both QGs and SFGs.
\section{X-ray Stacking Analysis}\label{sec:3}
\subsection{Stacking Procedure}\label{sec:3-1}
\par In this work, we use the Chandra COSMOS-Legacy survey data \citep{Civano2016} for the X-ray stacking. This survey is 4.6 Ms Chandra GO Program covering $2.2 \deg^2$ of the COSMOS field. The limiting depths are $2.2\times10^{-16}$, $1.5\times10^{-15}\ {\rm erg\ cm^{-2}\ s^{-1}}$ in the 0.5-2 keV, and 2-10 keV bands, respectively. For more detail, refer to \citet{Civano2016}.
\par In the stacking of X-ray images, we utilize the Chandra stacking tool  \citep[{\tt CSTACK} v.4.32\footnote{\url{http://lambic.astrosen.unam.mx/cstack/}},][]{Miyaji2008}. {\tt CSTACK} creates the stacked images at $0.5-2$ keV (soft band) and $2-8$ keV (hard band), separately. It first checks whether each object is located within $8.0\arcmin$ from the optical axis and not affected by resolved sources. Next, it generates a $30\arcsec\times30\arcsec$ cutout image of each object in the sample and sums up counts within the radius corresponding to $90\%$ of encircled counts fraction of the point spread function at each off-axis angle. In addition to the source count, it also estimates the background count from the outer region of images $>7\arcsec$ apart from the objects. From the estimated source and background counts, it derives a source count rate (CR) of each object. Finally, we obtain the exposure time-weighted mean count rate of the stacked sample. This procedure is conducted for both soft and hard bands. Hereafter, we use these values as the typical count rate of each subsample. We note that the physical scale is slightly different depending on the redshift. However, the source region radius is larger than the typical size of galaxies. In addition, the hot gas emission of galaxy groups or clusters will be negligible due to the rarity of massive halo.
\par The uncertainty of the mean count rate is derived via bootstrapping. {\tt CSTACK} reselects the sample from the original sample allowing duplication and reestimate the sample's mean count rate. This procedure is conducted 500 times, and the standard deviation of the mean count rate distribution is employed as $1\sigma$ uncertainty of the count rate.
\par  In this paper, we are focusing on the properties of typical QGs. Due to their high luminosity, the X-ray detected sources might affect the overall trend even though they are a small fraction in the entire sample. Therefore, all individually X-ray detected sources are removed from the sample. The galaxy catalog is cross-matched with the Chandra COSMOS-Legacy Survey source catalog, allowing the separation of $2.0\arcsec$. The fraction of the X-ray detected sources in each bin is summarized in Figure \ref{fig:3}. It varies from 0\% to 16\% dependent on the redshift and galaxy populations. In particular, galaxies with higher stellar mass tend to have a more significant fraction of X-ray detected sources for both QGs and SFGs at any redshift. All removed sources are point sources. Overall, we remove 668 and 1,261 X-ray-detected sources from our QG and SFG samples, respectively. 
\par
\begin{figure}
    \centering
    \includegraphics[width=7.3cm]{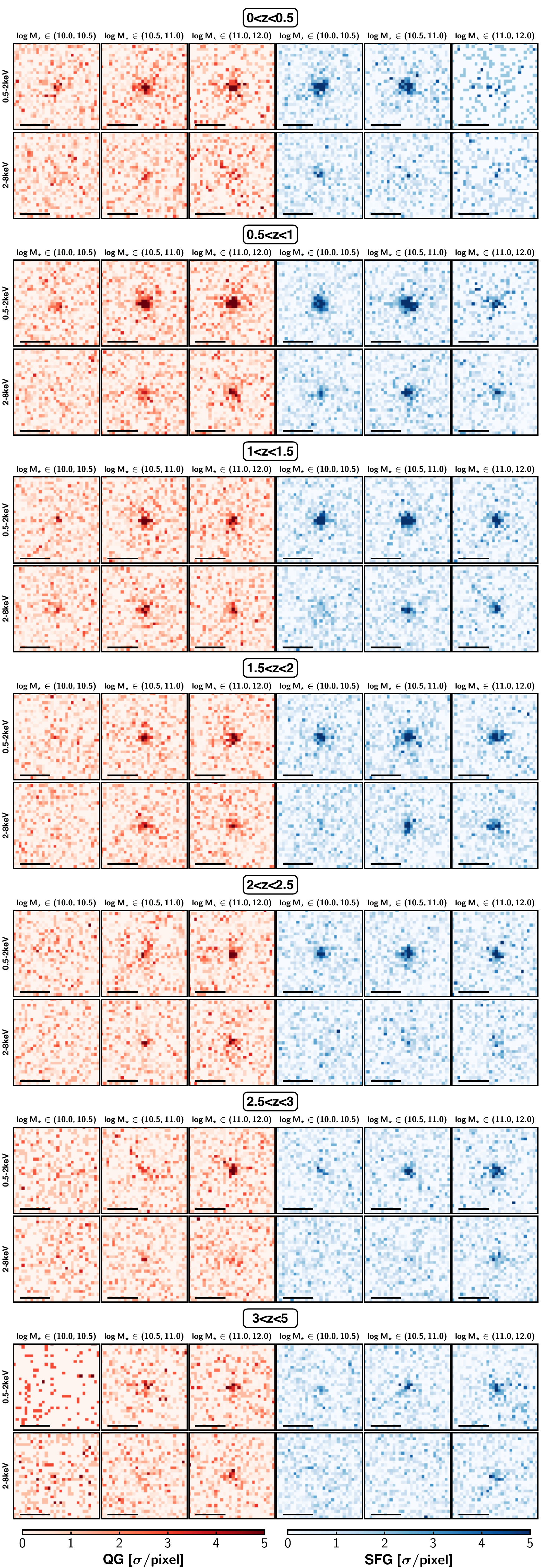}
    \caption{0.5-2 keV and 2-8 keV stacked images for QGs (red) and SFGs (blue) in each stellar mass and redshift bin. All images are $15\arcsec\times15\arcsec$ and shown with the same relative flux scale. The horizontal line corresponds to the scale of $5\arcsec$.}
    \label{fig:4}
\end{figure}

 The stacked images are summarized in Figure \ref{fig:4}. We can see the signal of the soft band for both QGs and SFGs in any redshift, at least with the $\log{(M_\star/M_\odot)}>10.5$. The signal-to-noise ratio is weaker in the hard band, but you can see the clear detection in some bins even at $z\sim3-5$. The derived count rate for each sample is summarized in Table \ref{tab:1}.
\subsection{Hardness Ratio and Spectral Evolution}\label{sec:3-3}
\par The hardness ratio (hereafter, referred to as HR), defined as ${\rm HR}=(H-S)/(H+S)$, is an indicator of the X-ray spectral shape, i.e., a combination of the photon index $\Gamma$ and the hydrogen column density ($N_H$). Here, $S\ {\rm and}\ H$ is the count rate of the soft and hard band, respectively. To discuss the spectral shape of our sample, we estimate the HR from the observed stacked count rates.
\par
\begin{figure*}
    \centering
    \includegraphics[width=16cm]{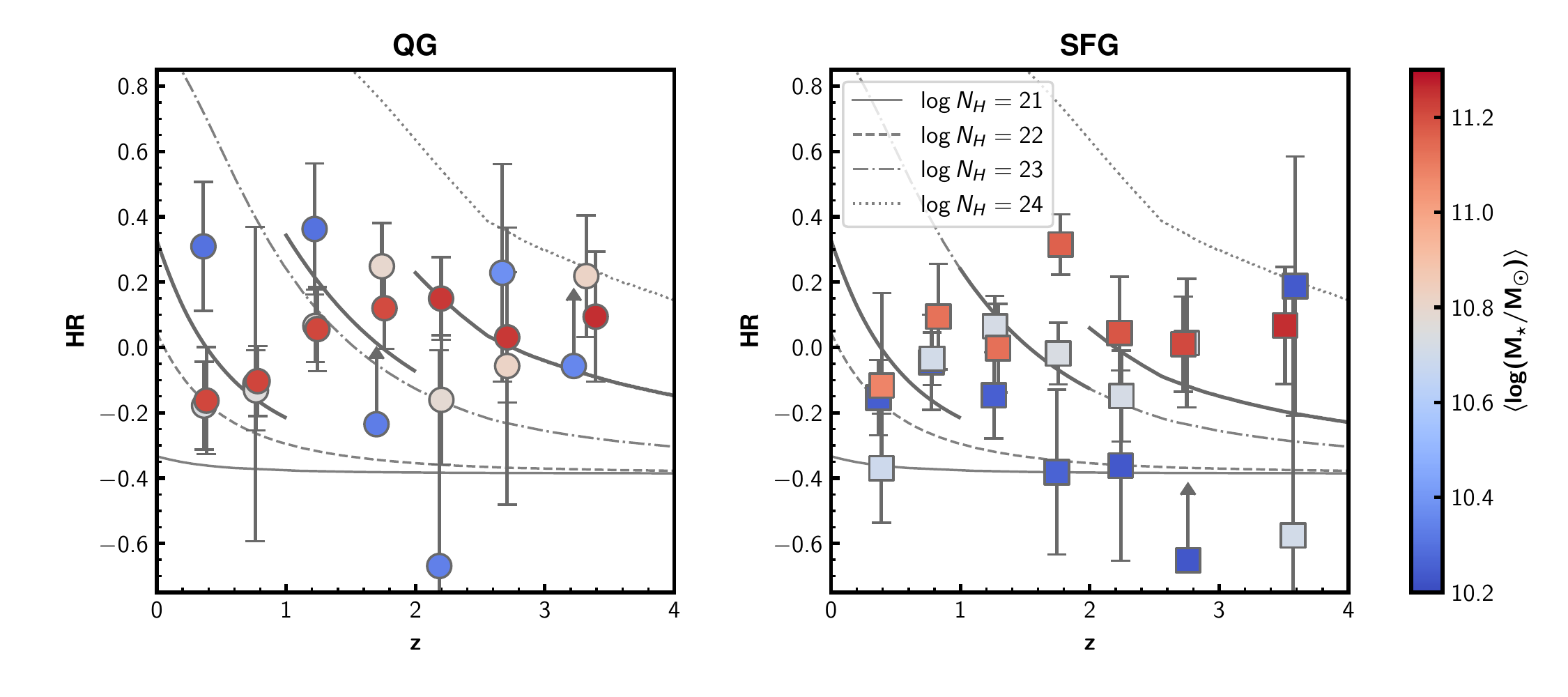}
    \caption{Hardness ratio of the stacked sample as a function of redshift. The left and right panels represent those of QGs and SFGs, respectively. The color of symbols indicates the average stellar mass of each sample. Bins of soft band flux with $S/N<2$ are shown in the $2\sigma$ lower limit (arrows). The thick gray lines show the expected HR value in the cases of the best-fit column densities. The gray solid, dashed, chain, and dotted lines show the hardness ratios of the power-law spectra with $\Gamma=1.8$ with the column densities of $\log{(N_H/{\rm cm^{-2}})}=21,\ 22,\ 23, {\rm and}\ 24$, respectively.}
    \label{fig:5}
\end{figure*}

 Figure \ref{fig:5} shows the HR as a function of the redshift of our stacked sample. Although their uncertainty is large, mainly due to the low sensitivity in the hard band, we can see a tentative trend that the HR increases with increasing redshift for both QGs and SFGs, whereas there is no significant dependency of the stellar mass. This trend is consistent with results in the literature  \citep[e.g.,][]{Fornasini2018,Carraro2020}. 
\par In order to derive the absorption corrected X-ray luminosity, it is essential to estimate the best column density value. Here, we compare the observed HR value to the model power-law spectrum absorbed by different neutral column densities derived by {\tt PIMMS}\footnote{\url{https://cxc.harvard.edu/toolkit/pimms.jsp}} tool \citep{Mukai1993} and the auxiliary response file of Cycle 14. In the model spectra, the photon index is fixed to be $\Gamma=1.8$, which is typical value of the AGN X-ray photon index \citep[e.g.,][]{Ricci2017,Marchesi2016a}. Its column density varies over $\log{(N_H/{\rm cm^{-2}})}=21-24$ with 0.1dex interval. The galactic absorption is also considered by assuming the column density as $N_H= 2.6\times10^{20}\ {\rm cm^{-2}}$ \citep{Kalberla2005}. In Figure \ref{fig:5}, we overplot the expected HR in the case of the column densities of $\log{(N_H/{\rm cm^{-2}})}=21,\ 22,\ 23,\ {\rm and}\ 24$. We conduct the least-squares fitting to derive the best column density for QGs and SFGs in three redshift cases ($0<z<1$, $1<z<2$, and $2<z<5$) by using all stellar mass and redshift bins with the detection of $S/N>2$ in either bands.
\par The estimated column densities are $\log{(N_H/{\rm cm^{-2}})}=22.3\ (22.3),\ 23.1\ (23.0),\ {\rm and}\ 23.5\ (23.3)$ for $0<z<1$, $1<z<2$, and $2<z<5$ of QGs (SFGs), respectively. The column density of QGs is higher than those of SFGs. Such dependency of the column density on the quiescence (i.e., sSFR) is similar to that reported in \citet{Fornasini2018}. They stack X-ray images of SFGs at $0.1<z<5$ selected from $UVJ$ diagram and report that the best column density is $\log{(N_H/{\rm cm^{-2}})}=22.2$ for galaxies with $\log{({\rm sSFR/yr^{-1}})}>-8.5$ and $\log{(N_H/{\rm cm^{-2}})}=22.0\ (23.0)$ for galaxies with $\log{({\rm sSFR/yr^{-1}})}\leq-8.5$ at $z<1.3\ (z>1.3)$ with the photon index of $\Gamma=1.4$. Hereafter, we employ our column densities to estimate the absorption corrected X-ray luminosities.
\subsection{Luminosity Estimation}
\par The source count rate is converted to the absorption-corrected flux via the conversion factor, which is estimated by the {\tt PIMMS} tool with the auxiliary response file of Cycle 14. Here, we assume the model spectrum of the power law with the slope with $\Gamma=1.8$ and correct for the galactic absorption and intrinsic absorption. The best column density value estimated in Section \ref{sec:3-3} is used in correcting the intrinsic absorption.
\par The absorption-corrected X-ray luminosity in the rest frame 2-10 keV is derived from the following equation. 

\begin{equation}
    L_{\rm 2-10keV} = \frac{4\pi d_L^2(10^{2-\Gamma}-2^{2-\Gamma})}{{(1+\bar{z})}^{2-\Gamma}(E_2^{2-\Gamma}-E_1^{2-\Gamma})}F_{X},
\end{equation}
where $F_{X}$ is the absorption-corrected flux in the observed $E_1-E_2$ keV band and the $d_L$ is the luminosity distance at the average redshift ($\bar{z}$) of the sample. Here, we use the soft band flux as $F_X$ (i.e., $E_1=0.5\ {\rm keV},\ E_2=2\ {\rm keV}$) since the Chandra effective area is larger in this band. Same as before, $\Gamma=1.8$ is assumed.  
\par
\begin{figure*}
    \centering
    \includegraphics[width=16cm]{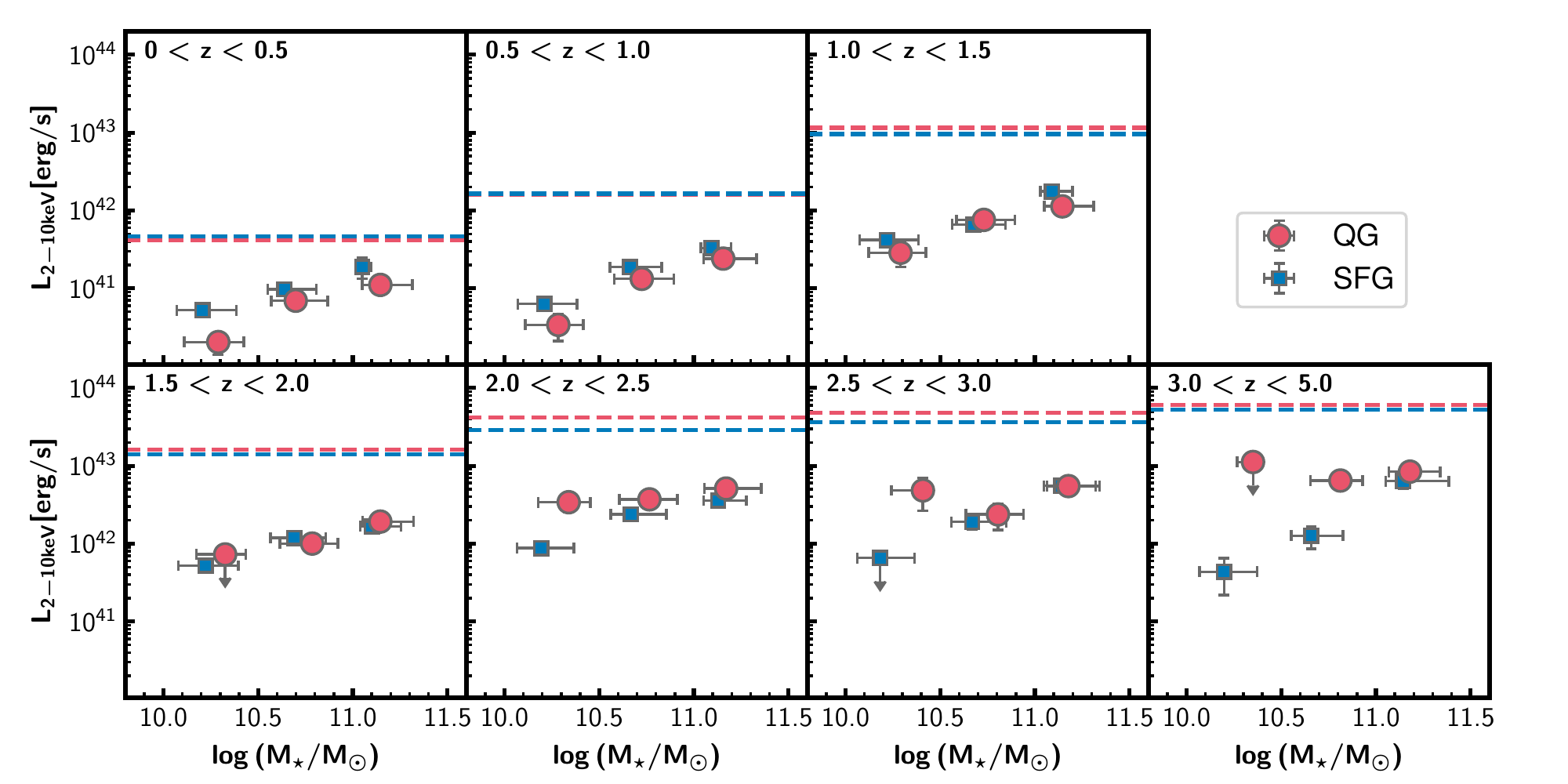}
\caption{Stacked X-ray luminosity of QGs (red circles) and SFGs (blue squares) as a function of redshift and stellar mass. Data points represent the median of the stellar mass and the average-stacked rest-frame 2-10keV luminosity. Their error bars on the vertical axis represent the $1\sigma$ uncertainty and those on the horizontal axis represent the 16th and 84th percentiles of the stellar mass distribution of each bin. The error bars of some bins are smaller than the size of the symbol. If the soft band flux has a signal to noise ratio less than two, the $2\sigma$ upper limit is plotted with the arrow. The limiting luminosity of the individual detection calculated from the limiting soft band flux and the best column density value for each redshift is shown in red and blue horizontal dashed lines for QGs and SFGs, respectively.}
    \label{fig:6}
\end{figure*}

 Figure \ref{fig:6} shows the luminosity as a function of stellar mass and redshift. From this figure, we make four points. Firstly, we successfully detect the signals of both QGs and SFGs and constrain their X-ray luminosity in all redshift bins. The most distant X-ray detection of individually non-detected QGs has been at $z\sim2$ \citep[][]{Olsen2013}. Thus, this study extends the X-ray detection of typical QGs up to $z\sim5$ for the first time. There are a few least massive bins ($10.0<\log{(M_\star/M_\odot)}<10.5$) at $z>1.5$ that do not yield a significant signal for both QGs and SFGs. These low mass QGs may be suffered from small number statistics as expected from the galaxy stellar mass functions \citep[e.g.,][]{Muzzin2013,Ilbert2013,Davidzon2017}. On the other hand, SFGs have a large sample number in that bin ($\sim5000$ objects), so this is likely to reflect the intrinsically small X-ray luminosity of SFGs. 
\par Secondly, we do not see any significant luminosity difference between QGs and SFGs at fixed redshift and stellar mass for most bins. However, the same luminosity between QGs and SFGs does not necessarily mean that the X-ray is due to the same mechanism since X-ray binaries (XRBs) contribute to the X-ray luminosity at different levels for different SFR and stellar masses. We discuss this point in Section \ref{sec:3-4}.
\par Thirdly, the luminosity increases with increasing stellar mass at the fixed redshift in both populations. \citet{Carraro2020} reports a similar trend for both QGs and SFGs with the average X-ray AGN luminosity of the sample including individually detected objects at $z<3.5$.
\par Lastly, the luminosity generally increases towards higher redshift for both QGs and SFGs. This trend can be the redshift evolution, but it can also be due to the selection bias because we focus on undetected objects in the source catalog, as mentioned in Section \ref{sec:3-1}. The limiting luminosity increases with increasing redshift, and thus the stacked sample covers a wider luminosity range for higher redshift bins. This can make the average luminosity possibly higher for higher redshift bins. For this reason, we hereafter do not discuss the redshift evolution of the value itself but only focus on the trend difference between QGs and SFGs at the same redshift.

\subsection{Contribution of XRBs and AGNs to X-ray Luminosity}\label{sec:3-4}
\par The X-ray emission in a galaxy comes from two main sources: XRBs and AGNs. The X-ray luminosity of the low-mass XRBs is correlated to the stellar mass, and that of the high-mass XRBs is correlated to the star formation activity \citep[e.g.,][]{Lehmer2010,Lehmer2016,Aird2017}. We estimate the contribution of XRBs to the observed X-ray luminosity to derive the AGN luminosity.
\par In this study, we use an empirical XRB scaling relation of galaxies at $z<4$ in the Chandra Deep Field South derived in \citet{Lehmer2016}. This relation is estimated in a similar way as ours and covers most of the redshift range that we are interested in this paper. There are other functional forms of the relation in the literature, and we show that there is only little effect of this assumption on our conclusions in Appendix \ref{sec:A}. The XRB luminosity, $L_{X,\mathrm{XRB}}$, is estimated from the average redshift, stellar mass, and SFR by the following relation:

\begin{multline}
    L_{X,\mathrm{XRB}}=10^{29.37\pm0.15}(1+z)^{2.0\pm0.6} M_{\star}\\
    +10^{39.28\pm0.05}(1+z)^{1.3\pm0.1} \mathrm{SFR}. \label{eq:2}
\end{multline}
\par
\begin{figure*}
    \centering
    \includegraphics[width=16cm]{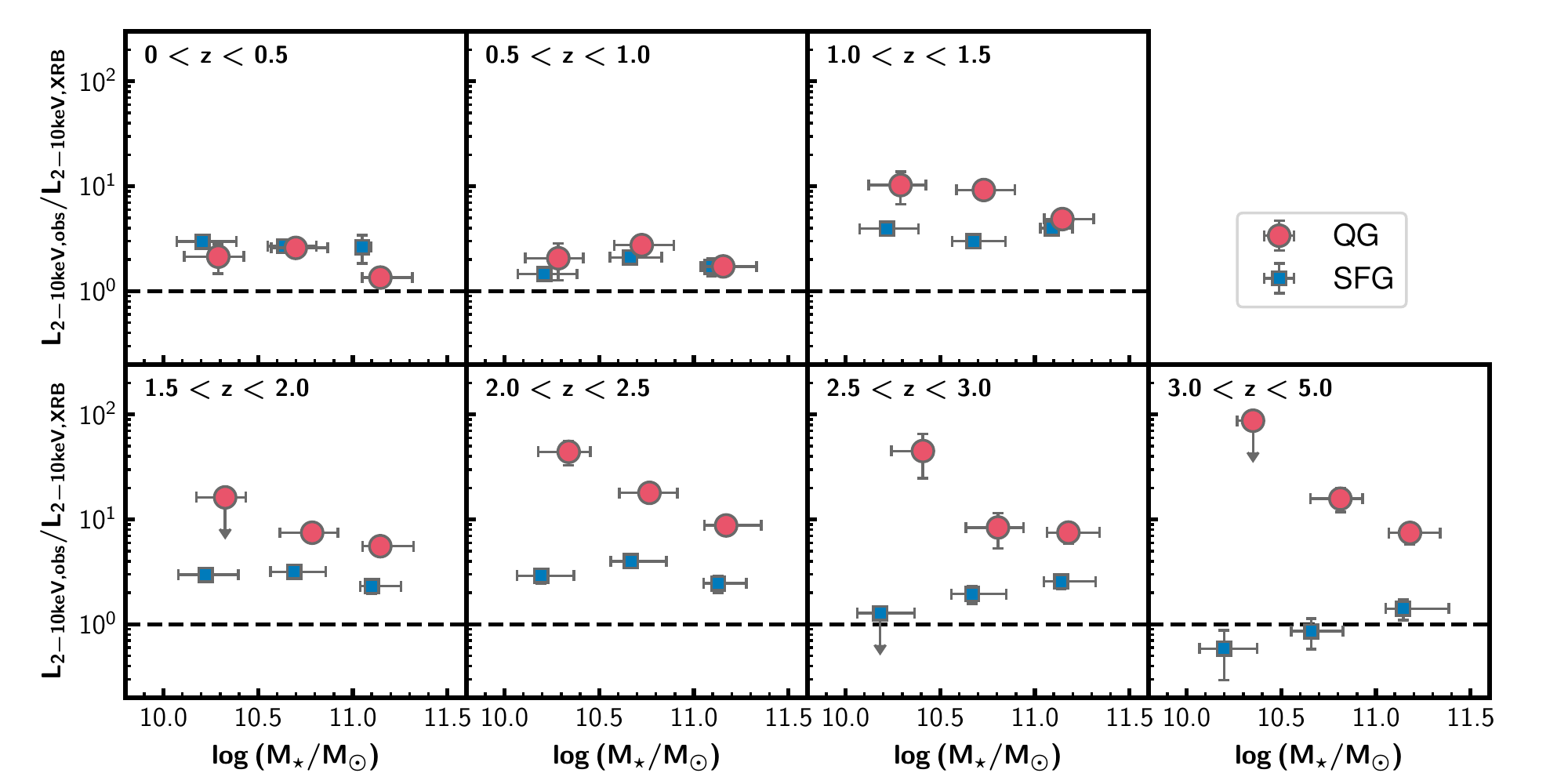}
    \caption{Ratio of the expected X-ray luminosity from XRBs and the observed X-ray luminosity as a function of redshift and stellar mass. The meanings of the symbols and the horizontal axis value of the data points are the same as in Figure \ref{fig:6}. We show the $2\sigma$ upper limit, if the soft band flux has a signal to noise ratio less than two.}
    \label{fig:7}
\end{figure*}

 Figure \ref{fig:7} shows the ratio of the XRB luminosity and the observed luminosity. The observed X-ray luminosity of SFGs is typically $\leq3$ times of the XRB luminosity at all redshift, which means that XRBs explain most of the observed X-ray luminosity of SFGs. On the other hand, the observed luminosity of QGs is higher than the expected XRB luminosity, especially by a factor of 5-50 at $z>1$. This indicates that the observed X-ray luminosity of QGs is much higher than the XRB luminosity and implies that AGNs are the dominant source of the X-ray emission of QGs.
\par
\begin{figure*}
    \centering
    \includegraphics[width=16cm]{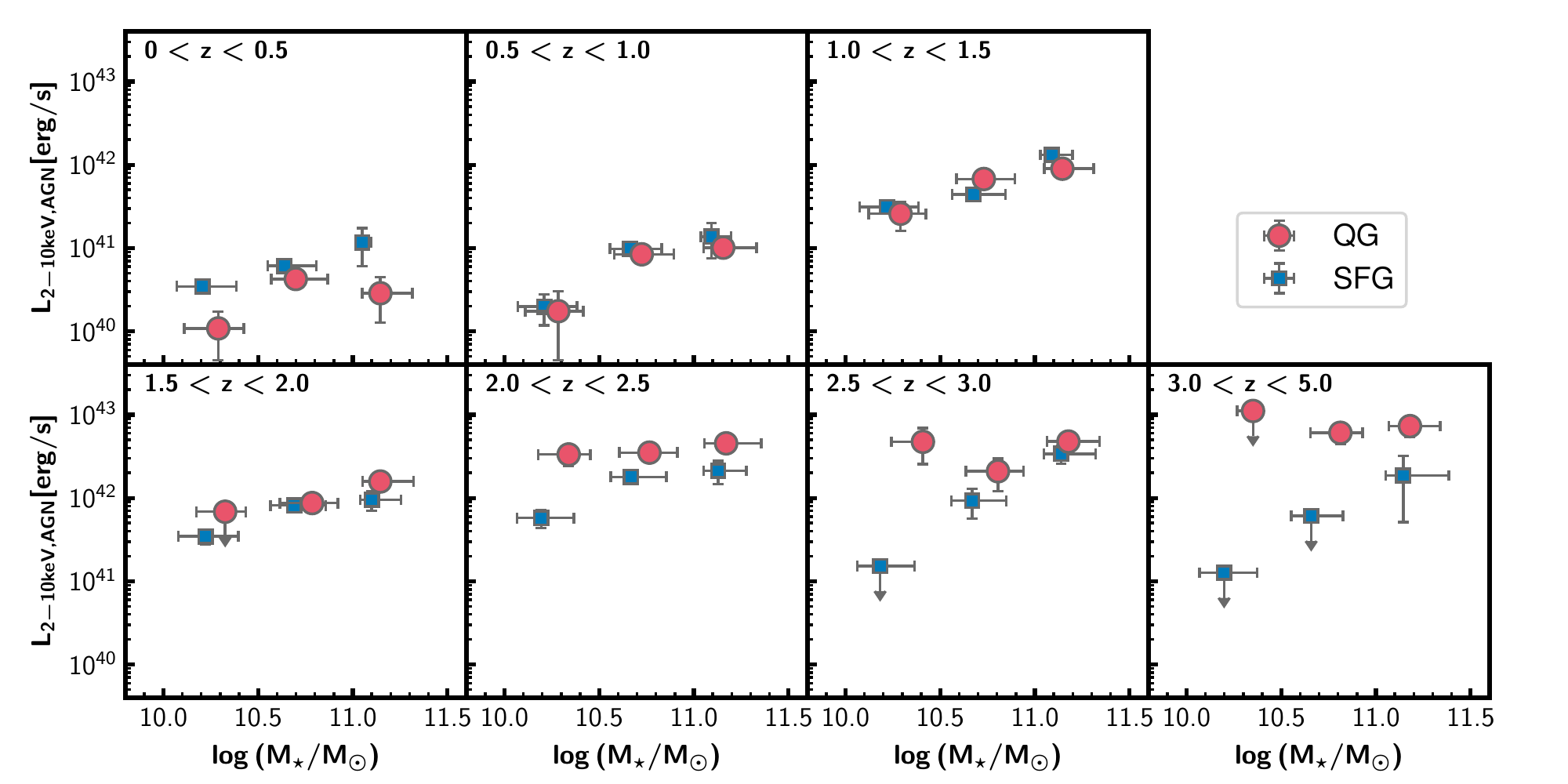}
    \caption{X-ray AGN luminosity as a function of redshift and stellar mass. The meanings of the symbols and the horizontal axis value of the data points are the same as in Figure \ref{fig:6}. If the luminosity has a signal to noise ratio less than two or the AGN luminosity is negative, the $2\sigma$ upper limit of the AGN luminosity is shown.}
    \label{fig:8}
\end{figure*}

 The excess of the observed X-ray luminosity to the expected XRB luminosity is interpreted as the AGN luminosity. Figure \ref{fig:8} shows the AGN luminosity in each redshift bin. Interestingly, at $z>1.5$, the AGN luminosity of QGs is higher than that of SFGs at any stellar mass bins. If we focus on bins at $z>1.5$ having the positive AGN luminosity for both populations, QGs have $\sim2.5$ times higher AGN luminosity than SFGs on average. Moreover, the difference between QGs and SFGs is the largest in the highest redshift bin, where QGs have $L_{X, \mathrm{AGN}}\sim(6-7)\times10^{42}$ erg/s, whereas the AGN luminosity of SFGs is lower than QGs, and it is consistent with zero for some subsample. It appears that QGs harbor more active AGNs than SFGs at these high redshifts, suggesting that AGNs may have played a role in quenching. On the other hand, at $z<1.5$, such enhancement of the AGN luminosity of QGs is not seen. Moreover, at the lowest redshift bin, SFGs have higher AGN luminosity than QGs. 
\par
\begin{figure}
    \centering
    \includegraphics[width=8.5cm]{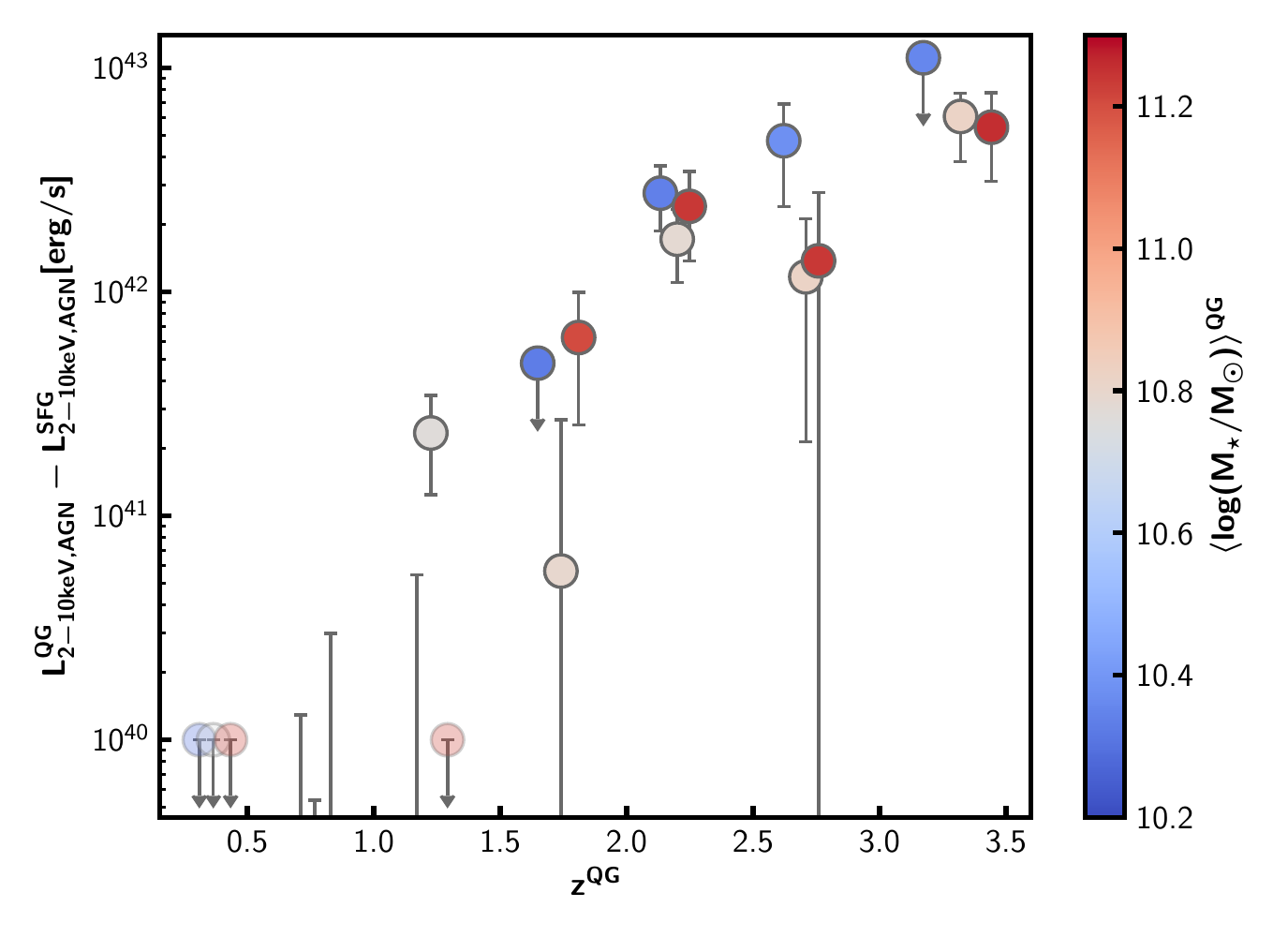}
    \caption{X-ray AGN luminosity excess of QGs to SFGs as a function of the redshift. The color of the marker shows the average stellar mass. If the AGN luminosity of QGs has only $2\sigma$ upper limit, $2\sigma$ upper limit of the excess is also shown in this figure. If the X-ray AGN luminosity excess is negative even considering the $1\sigma$ uncertainty, it is replaced by $1\times10^{40}$ erg/s only for illustrative purposes (shown in light color).}
    \label{fig:9}
\end{figure}

 This trend along the redshift is clearly seen in Figure \ref{fig:9}, which shows the excess of the X-ray AGN luminosity of QGs to that of SFGs. All bins at $z>1.5$ have positive values, whereas values of most of the bins at $z<1.5$ are consistent with zero or even negative. The observed trend has a significant implication for the quenching process. Before we discuss it, we examine another useful probe of AGN activity; radio emission.

\begin{deluxetable*}{cccccrrrrrr}
\tabletypesize{\scriptsize}
\tablecaption{Stacked X-ray properties of QGs and SFGs\label{tab:1}}
\tablehead{\colhead{ID}&\colhead{$z$ bin}&\colhead{$\log{M_\star}$ bin}&\colhead{$\langle z \rangle${\fontsize{3pt}{3pt}\selectfont \tablenotemark{a}}}&\colhead{$\langle\log{M_\star}\rangle${\fontsize{3pt}{3pt}\selectfont \tablenotemark{b}}}&\colhead{$\langle{\rm SFR}\rangle${\fontsize{3pt}{3pt}\selectfont \tablenotemark{c}}}&\colhead{soft CR}&\colhead{hard CR}&\colhead{HR{\fontsize{3pt}{3pt}\selectfont \tablenotemark{d}}}&\colhead{$L_X${\fontsize{3pt}{3pt}\selectfont \tablenotemark{e}}}&\colhead{$L_{X,{\rm AGN}}${\fontsize{3pt}{3pt}\selectfont \tablenotemark{e}}}\\
& &$\left(\log{M_\odot}\right)$& &$\left(\log{M_\odot}\right)$&$(M_\odot\ {\rm yr^{-1}})$ & ($\times 10^{-6}$ cts/s) & ($\times 10^{-6}$ cts/s) & &($\times 10^{41} {\rm erg/s}$) & ($\times 10^{41} {\rm erg/s}$)}
\startdata
\multicolumn{11}{c}{\textbf{QG}}\\
0  & (0.0,0.5) & (10.0,10.5) & 0.36 & $10.295 \pm 0.006$ & $  0.032 \pm 0.002 $ & $1.81 \pm 0.57$ & $ 3.4  \pm  1.1$ & $ 0.31\pm0.20$ & $ 0.204\pm 0.064$ & $ 0.109\pm 0.064$ \\
1  & (0.0,0.5) & (10.5,11.0) & 0.37 & $10.739 \pm 0.006$ & $  0.088 \pm 0.006 $ & $5.98 \pm 0.64$ & $ 4.2  \pm  1.1$ & $-0.18\pm0.13$ & $ 0.694\pm 0.074$ & $ 0.426\pm 0.074$ \\
2  & (0.0,0.5) & (11.0,12.0) & 0.38 & $11.214 \pm 0.014$ & $  0.224 \pm 0.025 $ & $ 8.8 \pm 1.2 $ & $ 6.3  \pm  1.9$ & $-0.16\pm0.16$ & $ 1.11\pm 0.16$ & $ 0.29\pm 0.16$ \\
3  & (0.5,1.0) & (10.0,10.5) & 0.76 & $10.292 \pm 0.003$ & $  0.128 \pm 0.003 $ & $0.77 \pm 0.29$ & $ 0.62 \pm 0.55$ & $-0.11\pm0.48$ & $ 0.34\pm 0.13$ & $ 0.17\pm 0.13$ \\
4  & (0.5,1.0) & (10.5,11.0) & 0.77 & $10.757 \pm 0.003$ & $  0.340 \pm 0.010 $ & $2.97 \pm 0.29$ & $ 2.28 \pm 0.52$ & $-0.13\pm0.12$ & $ 1.32\pm 0.13$ & $ 0.84\pm 0.13$ \\
5  & (0.5,1.0) & (11.0,12.0) & 0.78 & $11.218 \pm 0.006$ & $  0.650 \pm 0.032 $ & $5.24 \pm 0.49$ & $ 4.26 \pm 0.83$ & $-0.10\pm0.11$ & $ 2.41\pm 0.22$ & $ 1.01\pm 0.22$ \\
6  & (1.0,1.5) & (10.0,10.5) & 1.22 & $10.299 \pm 0.003$ & $  0.359 \pm 0.010 $ & $0.89 \pm 0.31$ & $ 1.90 \pm 0.59$ & $ 0.36\pm0.20$ & $ 2.87\pm 0.99$ & $ 2.59\pm 0.99$ \\
7  & (1.0,1.5) & (10.5,11.0) & 1.23 & $10.761 \pm 0.003$ & $  1.157 \pm 0.031 $ & $2.34 \pm 0.28$ & $ 2.67 \pm 0.51$ & $ 0.07\pm0.11$ & $ 7.58\pm 0.89$ & $ 6.76\pm 0.89$ \\
8  & (1.0,1.5) & (11.0,12.0) & 1.24 & $11.212 \pm 0.007$ & $  2.688 \pm 0.119 $ & $3.45 \pm 0.47$ & $ 3.86 \pm 0.84$ & $ 0.06\pm0.13$ & $11.3\pm 1.5$ & $ 9.0\pm 1.5$ \\
9  & (1.5,2.0) & (10.0,10.5) & 1.70 & $10.327 \pm 0.005$ & $  0.500 \pm 0.022 $ & $0.51 \pm 0.57$ & $ 2.2  \pm  1.1$ & $>-0.24       $ & $<7.3         $ & $<6.8         $ \\
10 & (1.5,2.0) & (10.5,11.0) & 1.74 & $10.794 \pm 0.004$ & $  1.333 \pm 0.044 $ & $2.21 \pm 0.41$ & $ 3.67 \pm 0.78$ & $ 0.25\pm0.13$ & $10.0\pm 1.9$ & $ 8.7\pm 1.9$ \\
11 & (1.5,2.0) & (11.0,12.0) & 1.76 & $11.202 \pm 0.007$ & $  3.090 \pm 0.158 $ & $4.18 \pm 0.60$ & $ 5.3  \pm  1.1$ & $ 0.12\pm0.12$ & $19.2\pm 2.8$ & $15.8\pm 2.8$ \\
12 & (2.0,2.5) & (10.0,10.5) & 2.18 & $10.335 \pm 0.007$ & $  1.976 \pm 0.095 $ & $2.93 \pm 0.76$ & $ 0.6  \pm  1.4$ & $-0.67\pm0.66$ & $34.1\pm 8.8$ & $33.3\pm 8.8$ \\
13 & (2.0,2.5) & (10.5,11.0) & 2.20 & $10.783 \pm 0.005$ & $  4.300 \pm 0.154 $ & $3.18 \pm 0.48$ & $ 2.30 \pm 0.87$ & $-0.16\pm0.20$ & $37.1\pm 5.6$ & $35.1\pm 5.6$ \\
14 & (2.0,2.5) & (11.0,12.0) & 2.20 & $11.236 \pm 0.010$ & $ 11.825 \pm 0.617 $ & $4.38 \pm 0.69$ & $ 5.9  \pm  1.2$ & $ 0.15\pm0.13$ & $51.1\pm 8.0$ & $45.3\pm 8.0$ \\
15 & (2.5,3.0) & (10.0,10.5) & 2.67 & $10.378 \pm 0.012$ & $  1.875 \pm 0.176 $ & $3.7  \pm  1.6$ & $ 5.8  \pm  3.1$ & $ 0.23\pm0.33$ & $48\pm22$ & $47\pm22$ \\
16 & (2.5,3.0) & (10.5,11.0) & 2.71 & $10.815 \pm 0.006$ & $  3.722 \pm 0.198 $ & $1.77 \pm 0.66$ & $ 1.6  \pm  1.2$ & $-0.06\pm0.42$ & $23.8\pm 8.8$ & $20.9\pm 8.8$ \\
17 & (2.5,3.0) & (11.0,12.0) & 2.71 & $11.236 \pm 0.011$ & $  8.423 \pm 0.660 $ & $4.09 \pm 0.85$ & $ 4.4  \pm  1.5$ & $ 0.03\pm0.20$ & $55\pm11$ & $48\pm11$ \\
18 & (3.0,5.0) & (10.0,10.5) & 3.22 & $10.351 \pm 0.018$ & $  1.507 \pm 0.327 $ & $1.4  \pm  2.8$ & $11.8  \pm  5.7$ & $>-0.06       $ & $<112        $ & $<111        $ \\
19 & (3.0,5.0) & (10.5,11.0) & 3.32 & $10.816 \pm 0.009$ & $  5.800 \pm 0.311 $ & $3.83 \pm 0.97$ & $ 6.0  \pm  1.8$ & $ 0.22\pm0.19$ & $65\pm16$ & $61\pm16$ \\
20 & (3.0,5.0) & (11.0,12.0) & 3.39 & $11.253 \pm 0.018$ & $ 13.819 \pm 1.350 $ & $4.9  \pm  1.1$ & $ 5.9  \pm  2.0$ & $ 0.09\pm0.20$ & $84\pm19$ & $73\pm19$ \\
\hline
\multicolumn{11}{c}{\textbf{SFG}}\\
0  & (0.0,0.5) & (10.0,10.5) & 0.37&$10.244 \pm 0.004$ & $  2.89 \pm  0.09 $ & $ 4.50 \pm 0.41$ & $ 3.30 \pm 0.72$ & $-0.15 \pm 0.12$ & $ 0.523 \pm  0.047$ & $ 0.35 \pm  0.048$ \\
1  & (0.0,0.5) & (10.5,11.0) & 0.39&$10.686 \pm 0.007$ & $  3.70 \pm  0.24 $ & $ 7.52 \pm 0.78$ & $ 3.5  \pm  1.3$ & $-0.37 \pm 0.17$ & $ 0.97 \pm  0.10$ & $ 0.61 \pm  0.10$ \\
2  & (0.0,0.5) & (11.0,12.0) & 0.39&$11.079 \pm 0.018$ & $  3.52 \pm  0.37 $ & $14.3  \pm  4.3$ & $11.3  \pm  5.5$ & $-0.12 \pm 0.28$ & $ 1.89 \pm  0.56$ & $ 1.17 \pm  0.56$ \\
3  & (0.5,1.0) & (10.0,10.5) & 0.78&$10.246 \pm 0.002$ & $  6.46 \pm  0.09 $ & $ 1.38 \pm 0.18$ & $ 1.26 \pm 0.33$ & $-0.05 \pm 0.15$ & $ 0.630 \pm 0.080$ & $ 0.198 \pm  0.080$ \\
4  & (0.5,1.0) & (10.5,11.0) & 0.78&$10.707 \pm 0.003$ & $ 10.56 \pm  0.26 $ & $ 4.05 \pm 0.30$ & $ 3.78 \pm 0.54$ & $-0.03 \pm 0.08$ & $ 1.87 \pm  0.14$ & $ 0.98 \pm  0.14$ \\
5  & (0.5,1.0) & (11.0,12.0) & 0.83&$11.121 \pm 0.008$ & $ 16.56 \pm  1.54 $ & $ 6.4  \pm  1.2$ & $ 7.7  \pm  2.1$ & $ 0.09 \pm 0.16$ & $ 3.30 \pm  0.62$ & $ 1.37 \pm  0.62$ \\
6  & (1.0,1.5) & (10.0,10.5) & 1.26&$10.251 \pm 0.002$ & $ 13.47 \pm  0.13 $ & $ 1.56 \pm 0.16$ & $ 1.16 \pm 0.29$ & $-0.15 \pm 0.13$ & $ 4.17 \pm  0.42$ & $ 3.11 \pm  0.42$ \\
7  & (1.0,1.5) & (10.5,11.0) & 1.27&$10.718 \pm 0.003$ & $ 24.58 \pm  0.38 $ & $ 2.46 \pm 0.24$ & $ 2.80 \pm 0.44$ & $ 0.06 \pm 0.09$ & $ 6.62 \pm  0.65$ & $ 4.41 \pm  0.65$ \\
8  & (1.0,1.5) & (11.0,12.0) & 1.29&$11.126 \pm 0.007$ & $ 41.58 \pm  2.49 $ & $ 6.40 \pm 0.87$ & $ 6.4  \pm  1.5$ & $-0.00 \pm 0.14$ & $17.6  \pm  2.4$ & $13.2 \pm  2.4$ \\
9  & (1.5,2.0) & (10.0,10.5) & 1.75&$10.258 \pm 0.002$ & $ 17.78 \pm  0.21 $ & $ 1.33 \pm 0.18$ & $ 0.60 \pm 0.34$ & $-0.38 \pm 0.25$ & $ 5.22 \pm  0.71$ & $ 3.46 \pm  0.71$ \\
10 & (1.5,2.0) & (10.5,11.0) & 1.76&$10.730 \pm 0.003$ & $ 33.61 \pm  0.62 $ & $ 3.00 \pm 0.26$ & $ 2.89 \pm 0.48$ & $-0.02 \pm 0.09$ & $11.9 \pm  1.0$ & $ 8.1 \pm  1.0$ \\
11 & (1.5,2.0) & (11.0,12.0) & 1.77&$11.160 \pm 0.007$ & $ 53.85 \pm  2.33 $ & $ 4.18 \pm 0.61$ & $ 8.0  \pm  1.2$ & $ 0.32 \pm 0.09$ & $16.7 \pm  2.5$ & $ 9.5 \pm  2.5$ \\
12 & (2.0,2.5) & (10.0,10.5) & 2.24&$10.234 \pm 0.002$ & $ 26.02 \pm  0.28 $ & $ 1.08 \pm 0.17$ & $ 0.51 \pm 0.33$ & $-0.36 \pm 0.29$ & $ 8.8 \pm  1.4$ & $ 5.8 \pm  1.4$ \\
13 & (2.0,2.5) & (10.5,11.0) & 2.24&$10.719 \pm 0.003$ & $ 46.00 \pm  1.04 $ & $ 2.93 \pm 0.32$ & $ 2.17 \pm 0.57$ & $-0.15 \pm 0.14$ & $23.8 \pm  2.6$ & $17.9 \pm  2.6$ \\
14 & (2.0,2.5) & (11.0,12.0) & 2.23&$11.183 \pm 0.009$ & $106.21 \pm  6.20 $ & $ 4.42 \pm 0.80$ & $ 4.8  \pm  1.4$ & $ 0.04 \pm 0.17$ & $35.8 \pm  6.5$ & $21.2 \pm  6.5$ \\
15 & (2.5,3.0) & (10.0,10.5) & 2.76&$10.229 \pm 0.002$ & $ 37.83 \pm  0.39 $ & $ 0.26 \pm 0.19$ & $ 0.50 \pm 0.37$ & $>-0.65         $ & $<6.6         $ & $<1.5        $ \\
16 & (2.5,3.0) & (10.5,11.0) & 2.75&$10.718 \pm 0.004$ & $ 66.34 \pm  1.72 $ & $ 1.87 \pm 0.36$ & $ 1.92 \pm 0.66$ & $ 0.01 \pm 0.20$ & $19.1 \pm  3.6$ & $ 9.3 \pm  3.6$ \\
17 & (2.5,3.0) & (11.0,12.0) & 2.72&$11.208 \pm 0.010$ & $134.00 \pm  8.46 $ & $ 5.51 \pm 0.81$ & $ 5.6  \pm  1.4$ & $ 0.01 \pm 0.15$ & $55.5 \pm  8.1$ & $34.0 \pm  8.2$ \\
18 & (3.0,5.0) & (10.0,10.5) & 3.59&$10.238 \pm 0.002$ & $ 41.98 \pm  0.30 $ & $ 0.29 \pm 0.14$ & $ 0.43 \pm 0.28$ & $ 0.19 \pm 0.40$ & $ 4.4 \pm  2.2$ & $-3.1 \pm  2.2$ \\
19 & (3.0,5.0) & (10.5,11.0) & 3.57&$10.703 \pm 0.003$ & $ 77.70 \pm  1.40 $ & $ 0.85 \pm 0.28$ & $ 0.23 \pm 0.52$ & $-0.58 \pm 0.76$ & $12.7 \pm  4.1$ & $-2.1 \pm  4.1$ \\
20 & (3.0,5.0) & (11.0,12.0) & 3.51&$11.256 \pm 0.016$ & $233.22 \pm 28.10 $ & $ 4.40 \pm 0.89$ & $ 5.0  \pm  1.5$ & $ 0.07 \pm 0.18$ & $64 \pm 13$ & $19 \pm 14$ \\
\enddata
\tablenotetext{a}{Average redshift of the sample}
\tablenotetext{b}{Average stellar mass of the sample}
\tablenotetext{b}{Average SFR of the sample}
\tablenotetext{d}{If the sample is detected with $S/N\leq2$ in the soft band, we show the $2\sigma$ lower limit.}
\tablenotetext{e}{If the sample is detected with $S/N\leq2$ in the soft band, we show the $2\sigma$ upper limit.}
\end{deluxetable*}

\section{Radio Stacking Analysis}\label{sec:4}
\subsection{Stacking Procedure, Flux Estimation}
\par  We use the imaging data at 3 GHz \citep[][]{Smolcic2017} and 1.4 GHz \citep[][]{Schinnerer2007} taken by Karl G. Jansky Very Large Array (VLA) to derive the average radio luminosity of the sample. The 3 GHz data is taken from VLA-COSMOS 3 GHz Large Project, and the total observation time is 384 hours, which leads to a median $1\sigma$ flux uncertainty of $2.3\ {\rm\mu Jy\ beam^{-1}}$. It covers $2\ {\rm deg^2}$ of the COSMOS field with the angular resolution of $0.75\arcsec\times0.75\arcsec$. The 1.4 GHz data is taken from VLA-COSMOS Large project, and the total observation time is 275 hours, which leads to a median $1\sigma$ flux uncertainty of $10.5(15)\ {\rm\mu Jy\ beam^{-1}}$ for  $1(2){\rm deg^2}$ of the COSMOS field. The angular resolution of 1.4 GHz data is $1.5\arcsec\times1.4\arcsec$.
\par The radio emission of all objects does not correlate with the X-ray emission, so some objects without X-ray detection can be detected in either 3 GHz or 1.4 GHz. Here, we aim to directly compare radio properties with X-ray properties. Therefore, we chose to use the same sample in the X-ray stacking, although a small fraction of the galaxies is individually detected in radio. The number of the radio-detected objects is obtained by cross-matching our galaxy sample with 3GHz and 1.4GHz catalogs, where the separation of 0.8\arcsec and 1\arcsec is allowed, respectively. As seen in Figure \ref{fig:2}, the fraction of the detected sources in the total sample ranges from 0\% to 45\% dependent on redshift, stellar mass, and galaxy population. In particular, this fraction is higher for SFGs with the higher stellar mass. This trend can be due to their high SFR. It is noted that the 1.4 GHz data does not cover a small part of the COSMOS field, and the objects there are removed. The fraction of removed objects is smaller than 1\% of the whole sample (186 objects), which is negligible enough not to alter our conclusions.
\par A similar method as the X-ray stacking is applied to the 3 GHz and 1.4 GHz images. We first generate cutout images of all galaxies with a size of $15\arcsec\times15\arcsec$. The systematic offset of the position in the radio image is corrected by using the best-fit linear relation in \citet{Smolcic2017}. We then derive the average of these images, which is referred to as the stacked image. In order to reduce the impact of nearby interlopers, $5\sigma$-clipping is applied when averaging. 
\par
\begin{figure}
    \centering
    \includegraphics[width=7.3cm]{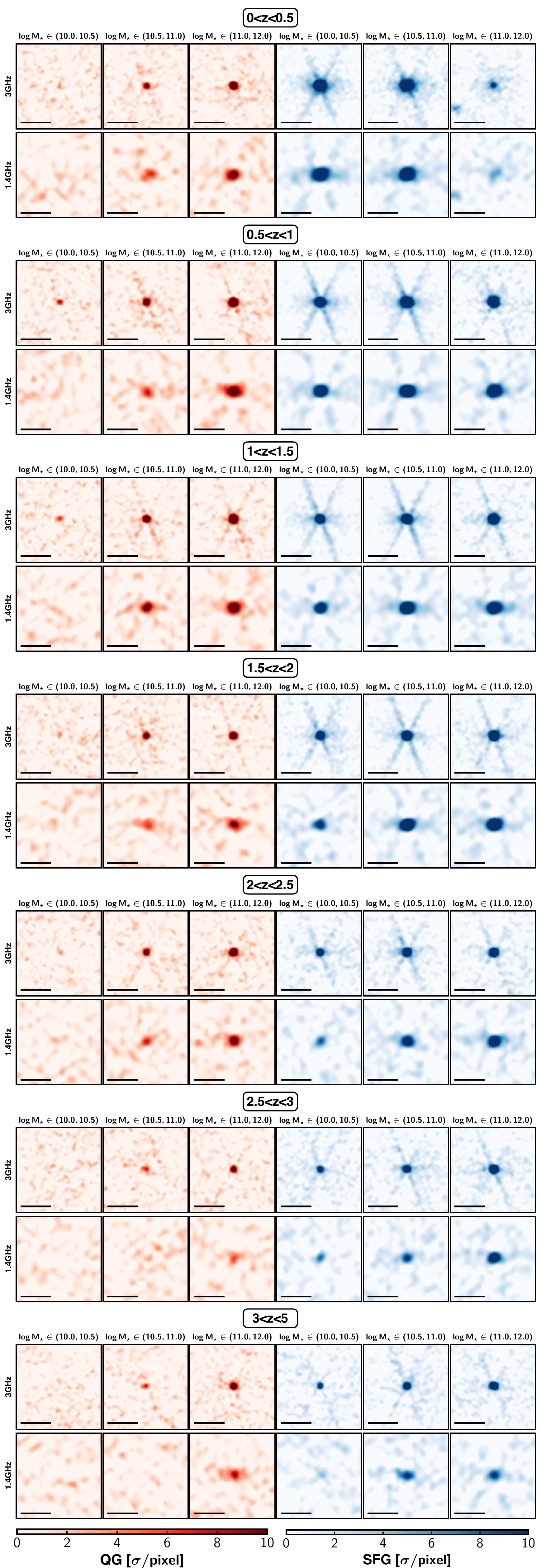}
    \caption{3 GHz and 1.4 GHz stacked image for QGs (red) and SFGs (blue) in each stellar mass and redshift bins. All images are $15\arcsec\times15\arcsec$ and shown with the same relative flux scale. The horizontal line corresponds to the scale of $5\arcsec$.}
    \label{fig:10}
\end{figure}

 The stacked image radio flux is estimated by fitting a 2D Gaussian profile to the central $8\arcsec\times8\arcsec$ of the images. Free parameters are its center position, sigma, and its amplitude. We use the integration of the best-fit 2D Gaussian profile as the total flux. The uncertainty of the flux is estimated using bootstrapping. It first reselects the same number of galaxies in the bin allowing for duplication. Averaging the images is then applied, and the flux is estimated by fitting the 2D Gaussian with the same center as the best fit of the original image. This trial is repeated 1000 times, and we take their standard deviation as the flux uncertainty. We show the stacked images of QGs and SFGs in Figure \ref{fig:10} and summarize the estimated fluxes in Table \ref{tab:2}. The 3GHz band signal is seen for subsamples with $\log{(M_\star/M_\odot)}>10.5$ of both QGs and SFGs at any redshift.

\subsection{Luminosity Estimation}
\par The spectral shape is assumed to be $S_\nu \propto \nu^\alpha$. The spectral index $\alpha$ is determined from the flux ratio of the observed 1.4 GHz and 3 GHz as follows:
\begin{equation}
    \alpha = \frac{\log{\left(\frac{F_{\rm 3GHz}}{F_{\rm 1.4GHz}}\right)}}{\log{\left(\frac{3}{1.4}\right)}},
\end{equation}
where $F_{\rm 3GHz},\ F_{\rm 1.4GHz}$ is the observed flux at 3 GHz and 1.4 GHz. Because the estimated flux at 1.4 GHz in many bins of QGs has large uncertainty ($S/N<2$), we assume that $\alpha=-0.75$ for all subsamples, which is an empirical value used for SFGs in the literature \citep[e.g.,][]{Delvecchio2020}. We note that the spectral slope does not significantly evolve with the redshift for SFGs, at least up to $z\sim2$ \citep{Magnelli2015}, supporting using the constant value at different redshift.
\par In this study, we discuss the radio luminosity at rest-frame 1.4 GHz, which is determined as follows:

\begin{equation}
    L_{\rm 1.4GHz}=\frac{4\pi d_{\rm L}^2}{(1+\bar{z})^{\alpha+1}}\left(\frac{1.4}{3}\right)^\alpha F_{\rm 3GHz},
\end{equation}
where $d_{\rm L}$ is the luminosity distance at that redshift and $F_{\rm 3GHz}$ is the observed flux at 3 GHz. The average redshift ($\bar{z}$) of galaxies in each bin is used in deriving the luminosity.
\par
\begin{figure*}
    \centering
    \includegraphics[width=16cm]{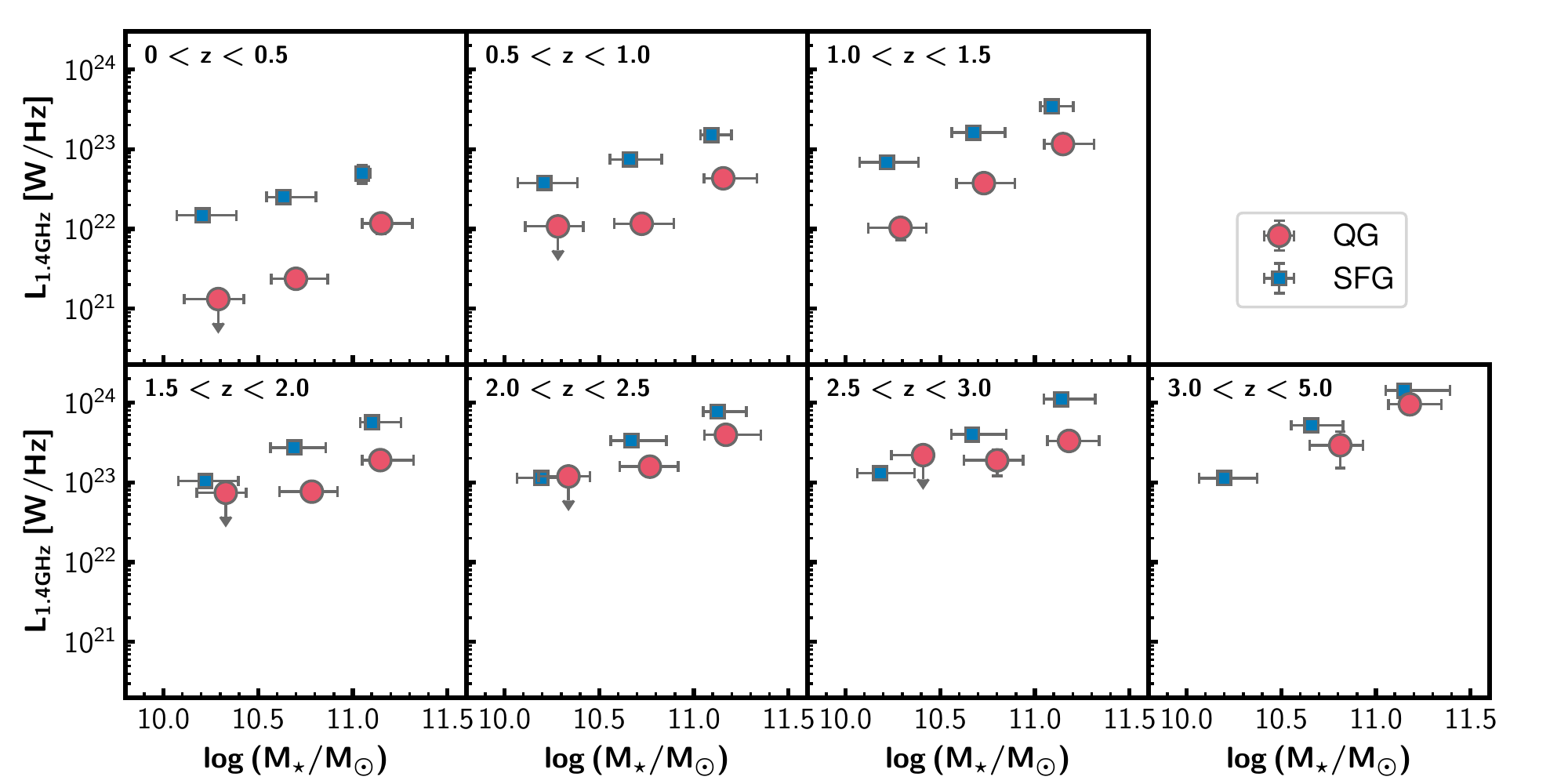}
    \caption{Stacked radio luminosity of QGs (red circles) and SFGs (blue squares) as a function of redshift and stellar mass. Data points represent the median of the stellar mass and the average-stacked rest-frame 1.4GHz luminosity. Their error bar on the vertical axis represent the $1\sigma$ uncertainty and those on the horizontal axis represent the 16th and 84th percentiles of the stellar mass distribution of each bin. The error bars of some bins are smaller than the size of the symbol. If the 3 GHz flux has a signal to noise ratio less than two, we show the $2\sigma$ upper limit. The least massive subsample of QGs at the highest redshift is missing due to the poor statistics.}
    \label{fig:11}
\end{figure*}

 Figure \ref{fig:11} shows the rest-frame 1.4 GHz luminosity of QGs and SFGs in each redshift bin. We successfully detect radio signals of QGs up to $z\sim5$ at least for $\log{(M_\star/M_\odot)}>10.5$. We find that SFGs systematically have a higher rest-frame 1.4 GHz luminosity than QGs at fixed stellar mass at all redshift, which is a different trend from X-ray analysis. In addition, the luminosity increases as the stellar mass increases for each galaxy population. As in the X-ray analysis, there are several origins of radio emission. To characterize the AGN activity, we need to account for the other origin, which is the subject of the following subsection.

\subsection{Rest-frame 1.4 GHz AGN Luminosity of QGs and SFGs}
\par The radio continuum of galaxies has mainly two origins. One is related to star formation activity, which is the synchrotron emission from supernovae and their remnants or the free-free emission from warm H{\sc ii} regions. The other is from AGNs. To investigate the contribution of AGN, we compare the observed luminosity and the expected contribution from star formation.
\par Similar to the X-ray analysis, we examine the contribution from star formation with the use of a known correlation between the total SFR and the rest-frame 1.4 GHz luminosity of SFGs. Here, we use an empirical relation derived in \citet{Delvecchio2020}, which is based on the stacking analysis of infrared and radio images for $NUVrJ$ selected SFGs at $0<z<4.5$:

\begin{multline}
    q_{SFR}(M_\star,z) = (2.743\pm0.034)\times(1+z)^{(-0.025\pm0.012)}\\
    -(0.234\pm0.017)\times(\log{(M_\star/M_\odot)}-10),\label{eq:5}
\end{multline}
where $q_{SFR}(M_\star,z)=\log{(L_{SFR}{\rm [W]}/(3.75\times10^{12}{\rm Hz}))}-\log{(L_{1.4 {\rm GHz}}\ {\rm [W\ Hz^{-1}]})}$, and $L_{SFR}$ is the luminosity equivalent to their SFR based on the correction factor in \citet{Kennicutt1998}.
\par There is a slight offset between the star formation main sequence of this work and those of \citet{Delvecchio2020}. We scale our SED-fitting based SFR to match that of \citet{Delvecchio2020} in estimating the expected luminosity from SFR from Equation \ref{eq:5}. To deal with the redshift bin difference between our study and \citet{Delvecchio2020}, we use the formulation of the redshift-dependent main sequence of \citet{Schreiber2015}, which is argued to be in good agreement with that of \citet{Delvecchio2020}. The ratio between our SFR and their expected SFR at fixed stellar mass and redshift is employed as the correction factor for SFR, which is in the range of $1.5-2.5$. We note that the results of the X-ray analysis do not significantly change whether or not these correction factors are applied.
\par
\begin{figure*}
    \centering
    \includegraphics[width=16cm]{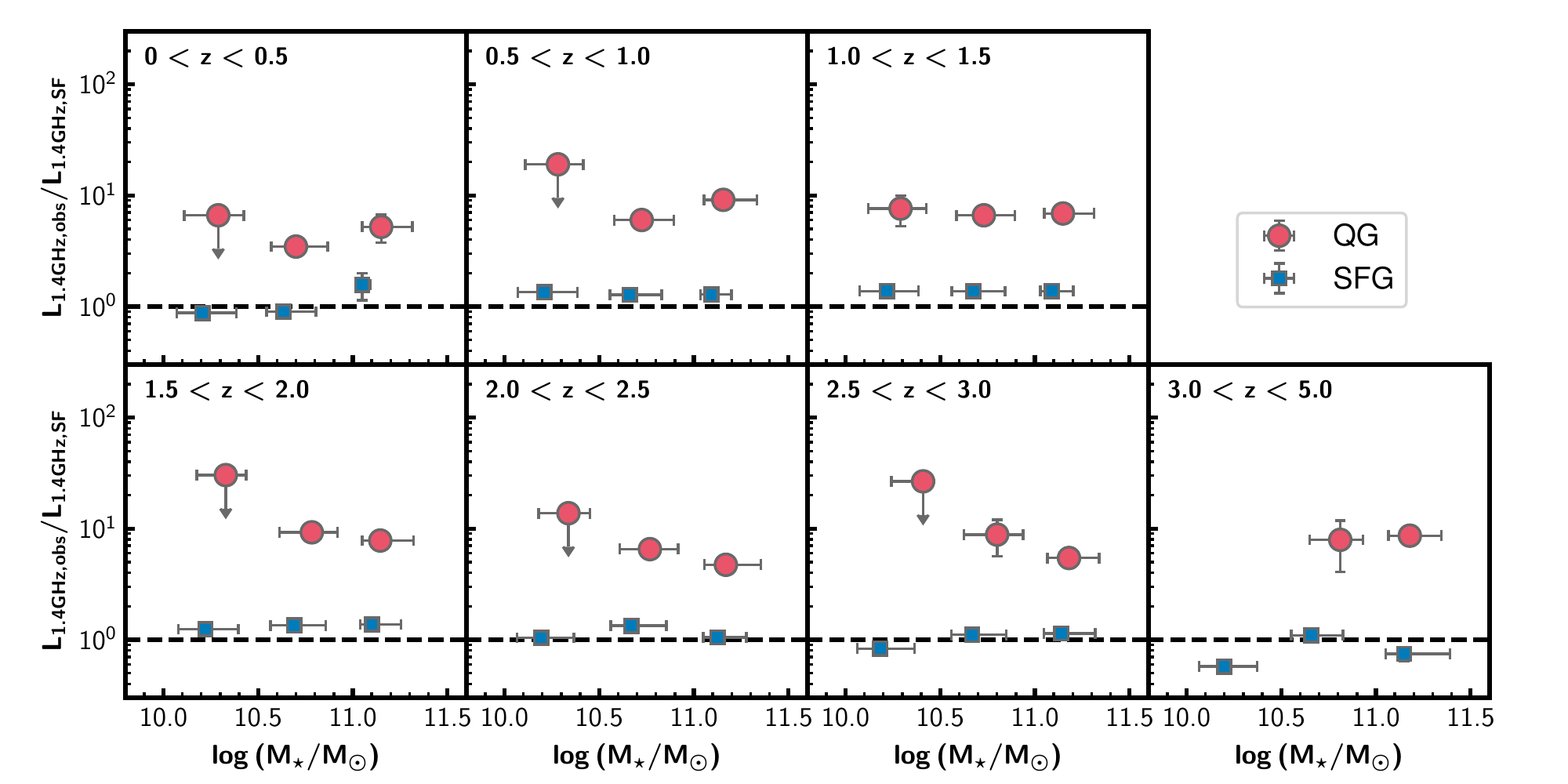}
    \caption{Ratio of observed radio luminosity and expected luminosity due to star formation for QGs and SFGs. The meanings of the symbols and the horizontal axis value of the data points are the same as in Figure \ref{fig:11}. The least massive subsample of QGs at the highest redshift is missing due to the poor statistics.}
    \label{fig:12}
\end{figure*}

 In Figure \ref{fig:12}, we compare the observed rest-frame 1.4 GHz luminosity with the luminosity from the star formation. QGs have 3-10 times higher luminosity than expected from their star formation at any redshift and stellar mass bins, whereas the observed 1.4 GHz luminosity of SFGs is comparable to that expected from their star formation. This high value suggests that the luminosity of QGs is mainly due to AGNs. This is fully consistent with our findings in Section \ref{sec:3}. Previous studies also report the enhancement of radio luminosity for color-selected QGs at $z<2$ \citep[e.g.,][]{Man2016,Gobat2017,Gobat2018,Magdis2021}, which lend further support to our results.
\par
\begin{figure*}
    \centering
    \includegraphics[width=16cm]{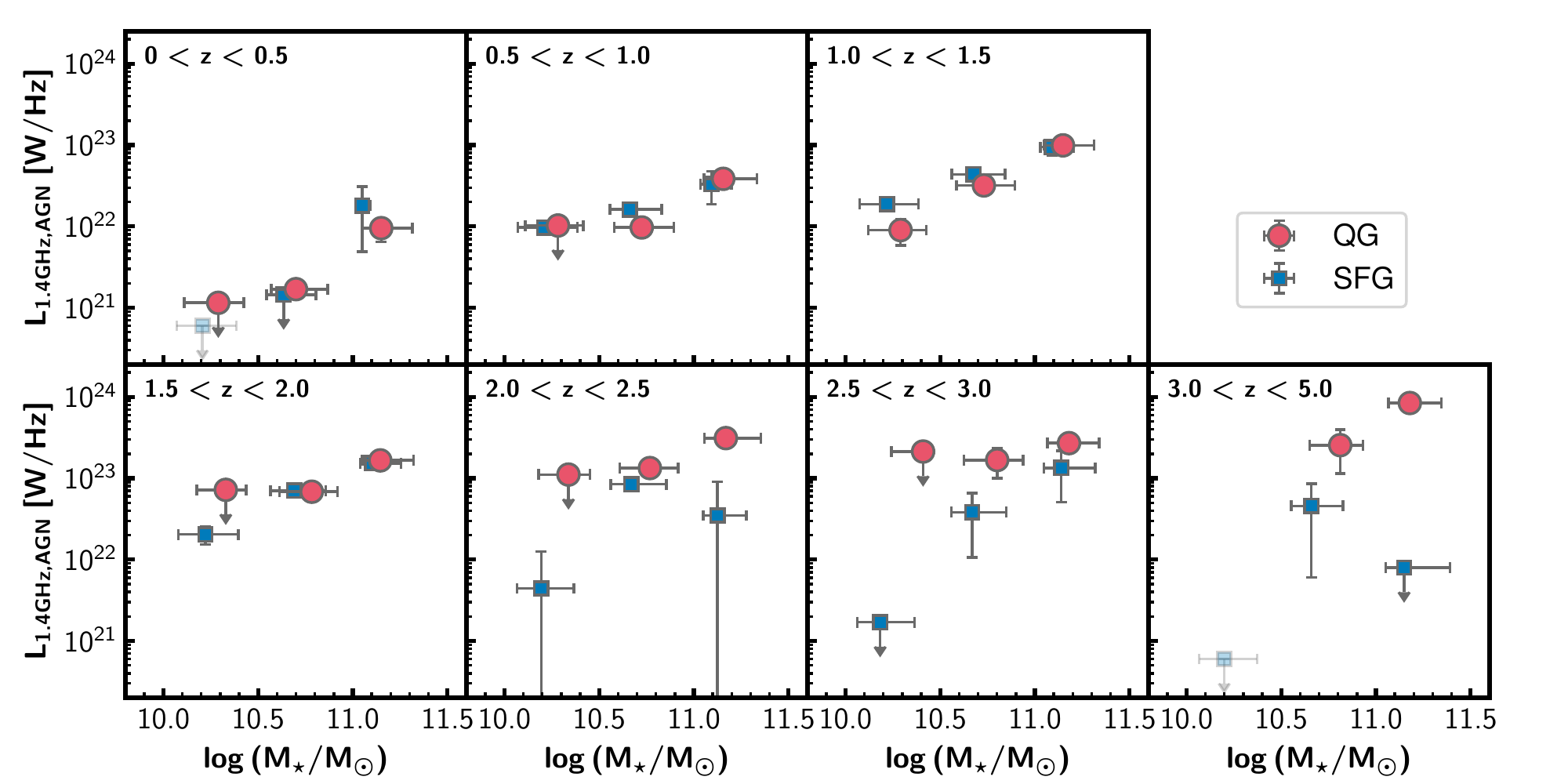}
    \caption{Radio AGN luminosity for QGs and SFGs as a function of redshift and stellar mass. The meanings of the symbols and the horizontal axis value of the data points are the same as in Figure \ref{fig:11}. The least massive subsample of QGs at the highest redshift is missing due to the poor statistics. If the luminosity has a signal to noise ratio less than two or the radio AGN luminosity is negative, the $2\sigma$ upper limit of the AGN luminosity is shown. If the upper limit is negative, it is replaced by $6\times10^{20}$ W/Hz only for illustrative purposes (shown in light color).}
    \label{fig:13}
\end{figure*}

\begin{figure}
    \centering
    \includegraphics[width=8.5cm]{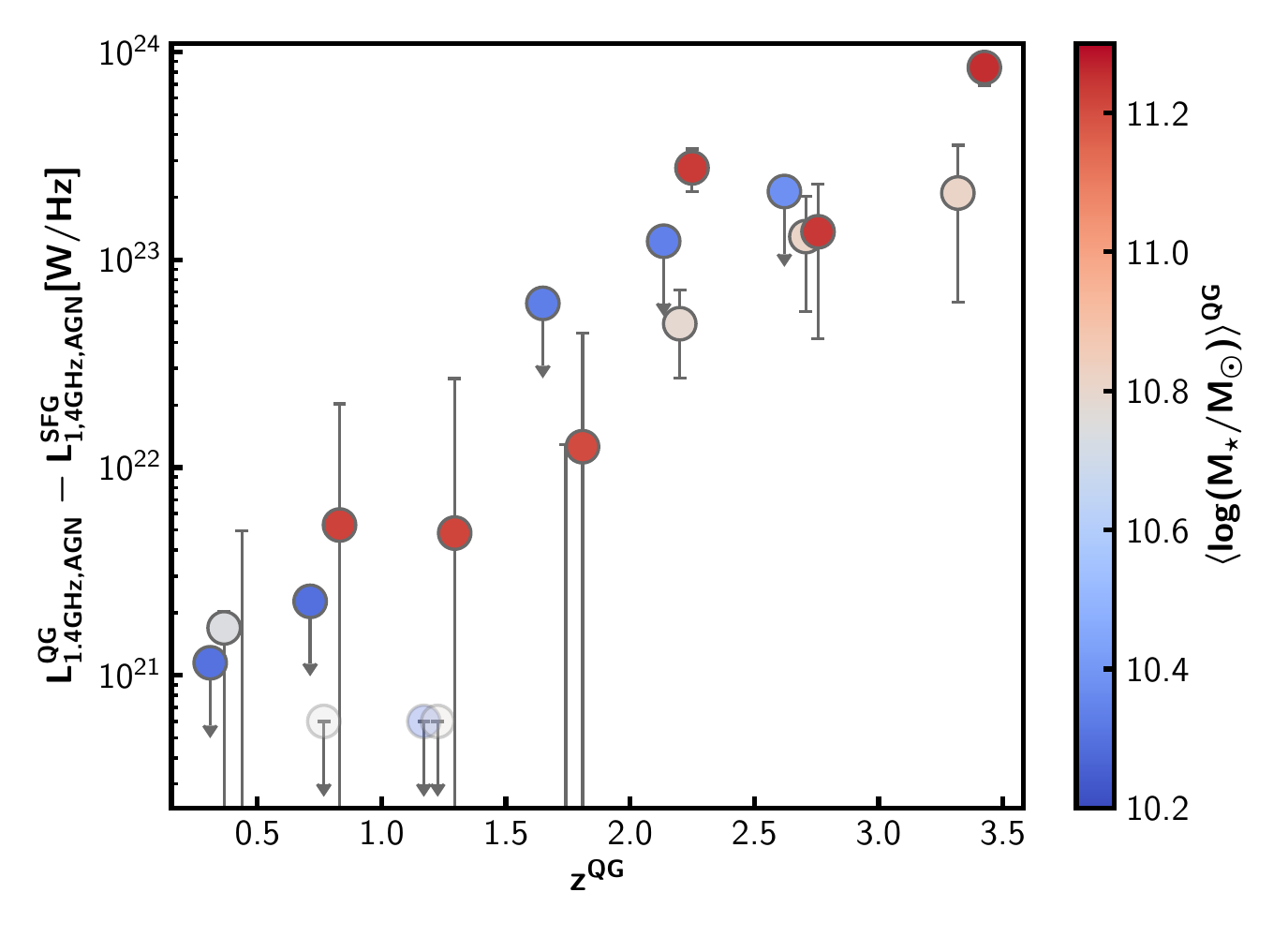}
    \caption{Radio AGN luminosity excess of QGs to SFGs as a function of the redshift. The color of the marker shows the average stellar mass. The least massive bin of QGs at the highest redshift is missing due to the poor statistics. If the radio AGN luminosity of QGs have only $2\sigma$ upper limit, $2\sigma$ upper limit is also shown in this figure. Also, if the $2\sigma$ upper limit of the radio AGN luminosity of SFGs is negative, we here show the radio AGN luminosity of QGs, assuming SFGs do not have AGN luminosity. If the radio AGN luminosity excess is negative even considering the $1\sigma$ uncertainty, it is replaced by $6\times10^{20}$ W/Hz only for illustrative purposes (shown in light color).}
    \label{fig:14}
\end{figure}

 We next estimate the AGN luminosity at rest-frame 1.4 GHz. Similar to the discussion of X-ray stacking, the rest-frame 1.4 GHz AGN luminosity is defined to be the excess of the observed luminosity to the expected luminosity from star formation. Figure \ref{fig:13} shows the evolution of the AGN luminosity of QGs and SFGs. Similar to the result in X-ray stacking, the radio AGN luminosity of QGs is higher at $z>1.5$ than those of SFGs and comparable at $z<1.5$. Figure \ref{fig:14}, which shows the excess of the radio AGN luminosity of QGs to that of SFGs, supports this observed trend. Once again, our findings here are entirely consistent with the X-ray analysis, demonstrating the robustness of our result. Now, we are in a position to discuss them in the context of galaxy quenching.
\begin{deluxetable}{crrrr}
\tabletypesize{\scriptsize}
\tablecaption{Stacked radio properties of QGs and SFGs\label{tab:2}}
\tablehead{\colhead{ID}&\colhead{3GHz Flux}&\colhead{1.4GHz Flux}&\colhead{$L_{\rm 1.4GHz}${\fontsize{3pt}{3pt}\selectfont \tablenotemark{a}}}&\colhead{$L_{\rm 1.4GHz,\ {\rm AGN}}${\fontsize{3pt}{3pt}\selectfont \tablenotemark{a}}}\\ 
&($\mu {\rm Jy}$)&($\mu {\rm Jy}$)&($\times 10^{22}\ {\rm W/Hz}$)&($\times 10^{22}\ {\rm W/Hz}$)}
\startdata
\multicolumn{5}{c}{\textbf{QG}}\\
0  & $ 0.81 \pm    0.51$ & $11    \pm    1564$ & $    <0.13          $ & $    <0.11          $ \\
1  & $ 3.17 \pm    0.43$ & $ 8.7 \pm    2.1$ & $     0.24 \pm  0.03$ & $     0.17 \pm  0.03$ \\
2  & $13.7 \pm    3.5$ & $25.3 \pm    6.5$ & $     1.18 \pm  0.30$ & $     0.95 \pm  0.30$ \\
3  & $ 1.01 \pm    0.81$ & $ 2.4 \pm    2.9$ & $    <1.1          $ & $    <1.0          $ \\
4  & $ 2.79 \pm    0.27$ & $ 8.6 \pm    2.6$ & $     1.16 \pm  0.11$ & $     0.97 \pm  0.11$ \\
5  & $ 9.9 \pm    1.0$ & $26.7 \pm    2.7$ & $     4.32 \pm  0.43$ & $     3.84 \pm  0.44$ \\
6  & $ 0.84 \pm    0.25$ & $ 6    \pm     934$ & $     1.04 \pm  0.31$ & $     0.90 \pm  0.31$ \\
7  & $ 3.00 \pm    0.19$ & $ 7.1 \pm    1.5$ & $     3.76 \pm  0.24$ & $     3.19 \pm  0.24$ \\
8  & $ 8.94 \pm    0.62$ & $19.2 \pm    2.1$ & $    11.60 \pm  0.81$ & $     9.92 \pm  0.81$ \\
9  & $ 0.85 \pm    0.96$ & $16.0 \pm    6.2$ & $    <7.4          $ & $    <7.2          $ \\
10 & $ 2.71 \pm    0.46$ & $ 7.9 \pm    1.7$ & $     7.7 \pm  1.3$ & $     6.8 \pm  1.3$ \\
11 & $ 6.57 \pm    0.56$ & $13.8 \pm    2.4$ & $    19.0 \pm  1.6$ & $    16.6 \pm  1.6$ \\
12 & $ 1.17 \pm    0.67$ & $ 2.7 \pm    1.8$ & $    <12          $ & $    <11          $ \\
13 & $ 3.26 \pm    0.36$ & $ 6.7 \pm    2.4$ & $    15.8 \pm  1.8$ & $    13.4 \pm  1.8$ \\
14 & $ 8.19 \pm    0.64$ & $16.4 \pm    2.4$ & $    39.5 \pm  3.1$ & $    31.2 \pm  3.1$ \\
15 & $ 0.6 \pm    1.2$ & $ 0 \pm     817$ & $   <22          $ & $   <21          $ \\
16 & $ 2.45 \pm    0.88$ & $11.8 \pm    6.6$ & $    18.9 \pm  6.8$ & $    16.8 \pm  6.8$ \\
17 & $ 4.30 \pm    0.58$ & $11.3 \pm    4.2$ & $    33.1 \pm  4.5$ & $    27.0 \pm  4.5$ \\
18 & $ 2    \pm    4726$ & $15    \pm     495$ & $<107349            $ & $ <107349           $ \\
19 & $ 2.4 \pm    1.2$ & $ 2 \pm     108$ & $    29 \pm 14$ & $    26 \pm 14$ \\
20 & $ 7.6 \pm    1.2$ & $26.4 \pm    6.1$ & $    95 \pm 15$ & $    84 \pm 15$ \\
\hline
\multicolumn{5}{c}{\textbf{SFG}}\\
0  & $19.69 \pm  0.77$ & $36.6 \pm  1.9$ & $  1.480 \pm  0.055$ & $ -0.211 \pm  0.076$ \\
1  & $29.0 \pm  1.3$ & $57.7 \pm  3.9$ & $  2.51 \pm  0.12$ & $ -0.28 \pm  0.21$ \\
2  & $   57 \pm    14$ & $92    \pm    21$ & $  5.0  \pm   1.3$ & $  1.8  \pm   1.3$ \\
3  & $ 8.80 \pm  0.18$ & $13.01 \pm  0.60$ & $  3.777 \pm  0.078$ & $  0.970 \pm  0.086$ \\
4  & $17.10 \pm  0.34$ & $30.1 \pm  1.0$ & $  7.46 \pm  0.14$ & $  1.61 \pm  0.20$ \\
5  & $30.0 \pm  1.9$ & $68.0 \pm  6.6$ & $ 15.09 \pm  0.97$ & $  3.3  \pm   1.4$ \\
6  & $ 5.14 \pm  0.14$ & $ 7.51 \pm  0.54$ & $  6.86 \pm  0.19$ & $  1.88 \pm  0.19$ \\
7  & $11.91 \pm  0.22$ & $20.93 \pm  0.95$ & $ 16.12 \pm  0.29$ & $  4.37 \pm  0.35$ \\
8  & $24.2 \pm  1.0$ & $52.4 \pm  4.1$ & $ 34.5  \pm   1.4$ & $  9.4  \pm   2.0$ \\
9  & $ 3.66 \pm  0.17$ & $ 4.24 \pm  0.49$ & $ 10.41 \pm  0.49$ & $  2.04 \pm  0.50$ \\
10 & $ 9.50 \pm  0.22$ & $17.28 \pm  0.92$ & $ 27.40 \pm  0.62$ & $ 7.03  \pm  0.73$ \\
11 & $19.17 \pm  0.71$ & $39.1 \pm  2.3$ & $ 56.6  \pm   2.1$ & $ 15.3  \pm   2.7$ \\
12 & $ 2.25 \pm  0.16$ & $ 2.81 \pm  0.44$ & $ 11.34  \pm  0.81$ & $  0.44 \pm  0.82$ \\
13 & $ 6.62 \pm  0.25$ & $12.5 \pm  1.2$ & $ 33.5  \pm   1.2$ & $  8.4  \pm   1.4$ \\
14 & $15.49 \pm  0.74$ & $35.8 \pm  3.8$ & $ 77.2  \pm   3.7$ & $  3.5  \pm   5.6$ \\
15 & $ 1.63 \pm  0.17$ & $ 1.90 \pm  0.37$ & $ 13.1  \pm   1.4$ & $ -2.7  \pm   1.4$ \\
16 & $ 5.01 \pm  0.32$ & $ 9.0 \pm  1.7$ & $ 39.9  \pm   2.6$ & $  3.8  \pm   2.8$ \\
17 & $14.15 \pm  0.75$ & $23.6 \pm  2.6$ & $110.5  \pm   5.8$ & $ 13.4  \pm   8.3$ \\
18 & $ 0.791 \pm  0.087$ & $ 1.27 \pm  0.59$ & $ 11.3  \pm   1.2$ & $ -8.4  \pm   1.3$ \\
19 & $ 3.66 \pm  0.27$ & $ 6.94 \pm  0.98$ & $ 52.0  \pm   3.9$ & $  4.6  \pm   4.0$ \\
20 & $10.42 \pm  0.76$ & $18.9 \pm  3.0$ & $142    \pm    10$ & $-49    \pm    25$ \\
\enddata
\tablenotetext{a}{If the 3GHz flux is detected with $S/N<2$, we show the upper limit of the estimated luminosity.}
\end{deluxetable}

\section{Discussion}\label{sec:5}
So far, we find that QGs have X-ray and radio emissions which are dominantly from AGNs, and their luminosity is higher than SFGs with the same stellar mass at $z>1.5$. Here, we will discuss implications for the galaxy quenching.
\subsection{AGN Activity and Galaxy Populations}\label{sec:5-1}
\par
\begin{figure}
    \includegraphics[width=8.7cm]{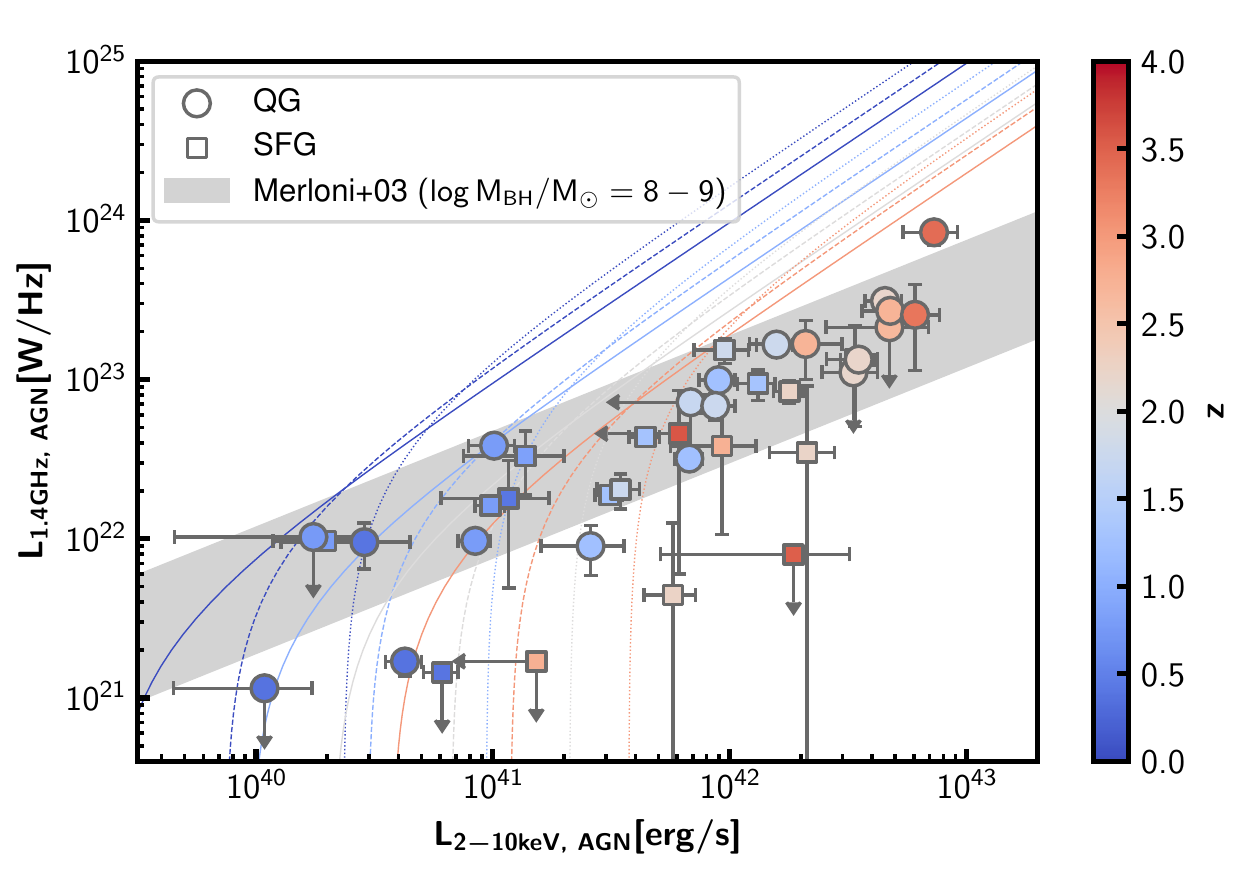}
    \caption{Comparison between the X-ray and radio AGN luminosity of QGs (circles) and SFGs (squares). The color of each marker shows the average redshift of each bin. Same as in Figure \ref{fig:8} and Figure \ref{fig:13}, if the observed luminosity has a signal to noise ratio of less than two or the AGN luminosity is negative, the $2\sigma$ upper limit is shown. Negative upper limits in radio are not shown, assuming that they do not contain AGNs. Following Figure \ref{fig:13}, the least massive bin at $3.0<z<5.0$ is not shown due to the poor statistic in the radio stacking. An expected relation of the X-ray and radio luminosity for star formation from Equation \ref{eq:2} and \ref{eq:5} is shown for bins with $\log{(M_\star/M_\odot)}=10.0,\ 10.5,\ {\rm and}\ 11.0$ in solid, dashed, and dotted lines, at $z=0,\ 1,\ 2,\ {\rm and\ }3$, following the same color map as symbols. The local relation with black hole mass of $\log{(M_{\rm BH}/M_\odot)}=8-9$ from \citet{Merloni2003} after correcting to the unit in this study is shown in gray hatched region.}\label{fig:15}
\end{figure}

 We first compare the X-ray and radio AGN luminosity in Figure \ref{fig:15}. The AGN luminosity estimated in two different wavelengths is correlated well with the same relation for both QGs and SFGs. We calculate Spearman's rank correlation coefficients $\rho$ between the X-ray and radio AGN luminosity bins which have significant positive detection in both wavelengths. The Spearman's $\rho$ value is estimated as $\rho=0.94$ with $P=2\times10^{-7}$ and $\rho=0.77$ with $P=0.001$ for QGs and SFGs, respectively. These values support the strong correlation between the X-ray and radio luminosity. Given the redshift and the stellar mass, the relation of the X-ray and the radio luminosity for the star formation is derived from Equation \ref{eq:2} and Equation \ref{eq:5}. Our result has lower radio luminosity than the star formation relation at the fixed X-ray luminosity and redshift, again reinforcing that the emission does not originate from star formation. Moreover, the slope of the relation is similar to that of ``fundamental plane of the black hole activity" with $L_{\rm radio}\propto L_{X}^{0.6}$ in the local universe \citep{Merloni2003}. In particular, the observed relation has a same amplitude to their local relation with the black hole mass of $\log{(M_{\rm BH}/M_\odot)}=8-9$, after converting the rest-frame 5GHz luminosity to the rest-frame 1.4 GHz luminosity with $\alpha=-0.75$. This suggests that this emission is likely from AGN with the possible disc-jet connection \citep[e.g.,][]{Chatterjee2009}. We show for the first time that the low luminosity AGNs of QGs are located along with the local relation even up to $z\sim5$. As seen in Figure \ref{fig:13} and Figure \ref{fig:15}, the higher stellar mass bins tend to have higher radio AGN luminosity at fixed redshift, whereas they have the same X-ray AGN luminosity. According to the relation of \citet{Merloni2003}, this might prove the higher black hole mass for more massive galaxies, as seen in the local universe \citep[e.g.,][]{Kormendy2013}.
\par In addition, the X-ray AGN luminosity can be converted to the black hole accretion rate (BHAR) via a bolometric correction and an assumption of the accretion efficiency. The average bolometric luminosity is derived from the average X-ray AGN luminosity by using the conversion factor $k_{\rm bol}$  ($k_{\rm bol} \equiv L_{\rm bol}/L_X$) used in \citet{Yang2018}, which is a modified version of \citet{Lusso2012} for subsample with $\log{L_X[{\rm erg/s}]}>42.4$. For subsample with $\log{L_X[{\rm erg/s}]}<42.4$, $k_{\rm bol}=16$ is used as in \citet{She2017}. The average BHAR, $\langle {\rm BHAR} \rangle$, is then derived from the average bolometric luminosity as follows:

\begin{equation}
    \langle{\rm BHAR}\rangle(M_\star,z) = \frac{(1-\epsilon)\times \langle L_{\rm bol}\rangle(M_\star,z)}{\epsilon c^2},
\end{equation}
where, $c$ is the speed of the light and $\epsilon$ is the efficiency of the mass conversion. There is an uncertainty in the value of $\epsilon$, but following previous studies \citep[e.g.,][]{Carraro2020,DEugenio2020}, we assume $\epsilon=0.1$.
\par
\begin{figure*}
    \centering
    \includegraphics[width=14cm]{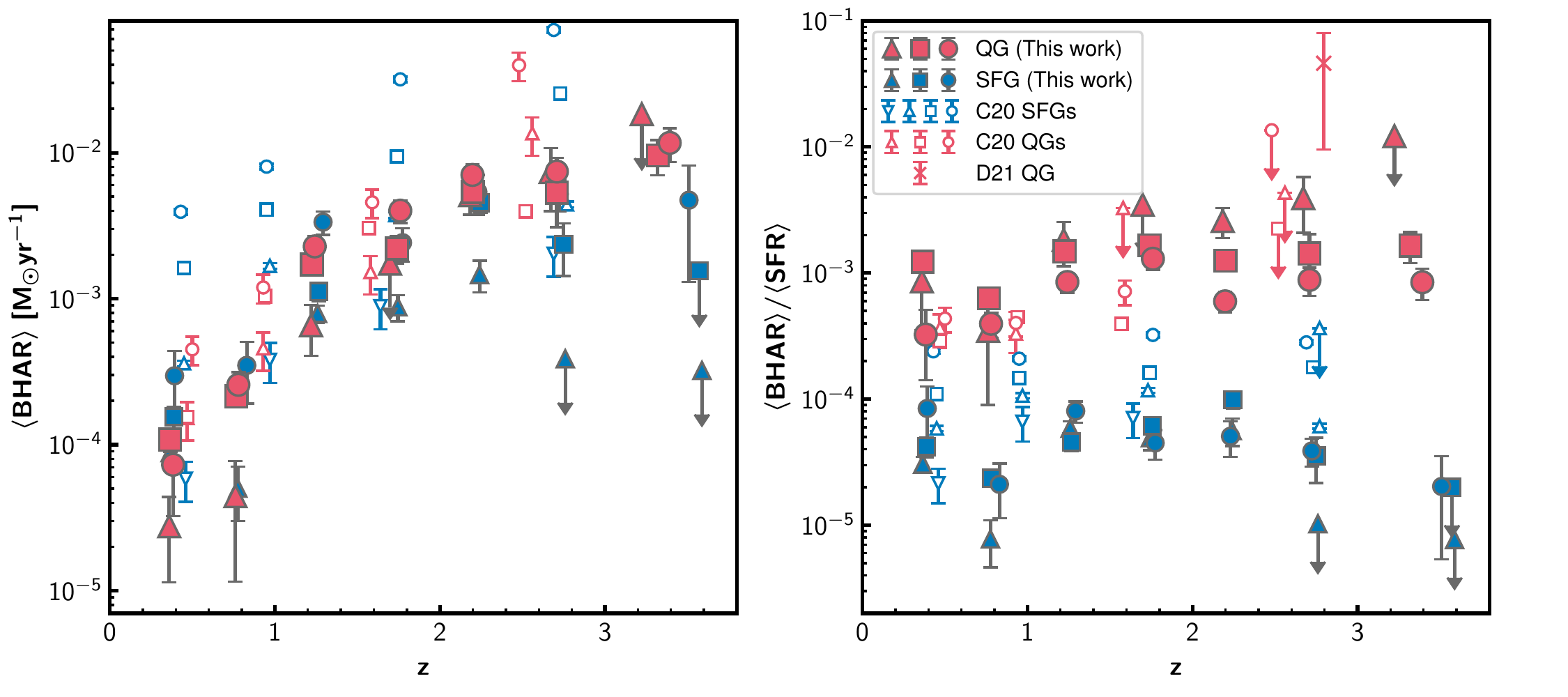}
    \caption{{\bf Left panel:} Redshift evolution of BHAR. The red and blue filled symbols show the QGs and SFGs, respectively. The triangles, squares, and circles represent bins of $10.0<\log{(M_\star/M_\odot)}<10.5$, $10.5<\log{(M_\star/M_\odot)}<11.0$, and $11.0<\log{(M_\star/M_\odot)}<12.0$ for each population, respectively. The red and blue open symbols show the BHARs of \citet{Carraro2020}. Their stellar mass bins are shown in different symbols. {\bf Right panel:} Ratio of BHAR and SFR. The triangles, squares, and circles represent same as in the left panel. The red cross shows the average ratio of QGs in \citet{DEugenio2020}, which include individually X-ray detection.}
    \label{fig:16}
\end{figure*}

 The left panel of Figure \ref{fig:16} shows the average BHAR evolution of QGs and SFGs. We estimate BHARs of QGs at $z>3$ for the first time. As suggested from the X-ray AGN luminosity, BHARs of QGs are higher than those of SFGs at $z>1.5$. The BHAR from \citet{Carraro2020} is also overplotted and larger than ours. This is likely because \citet{Carraro2020} included individually detected X-ray sources. Interestingly, the higher BHAR for QGs is not seen in their study, which is different from our observed trend. This may indicate that such enhanced AGN activity of QGs only occurs at relatively low flux levels, and the inclusion of individually detected sources may hide this trend. It can also be due to the different classification methods of QGs ($NUVrJ$ diagram) from this study. 
\par The right panel of Figure \ref{fig:16} shows the ratio of BHAR and SFR of this study and those from the literature. We find that QGs have larger ratios than SFGs. This is because the BHAR of QGs is higher than those of SFGs and the SFR of QGs is lower than those of SFGs. This trend implies that QGs do not only have higher AGN activity than SFGs but also have a higher rate of gas accretion towards the nuclei compared to the gas consumption by star formation than SFGs. In addition, the ${\rm \langle BHAR\rangle/\langle SFR\rangle}$ ratio of QGs is almost constant (${\rm \langle BHAR\rangle/\langle SFR\rangle}\sim10^{-3}$) regardless of redshift. This might suggest that the growth rate of stellar and black hole mass ratio does not have significant redshift evolution. We note that our work only focuses on nondetected objects, and thus this ratio is lower than other studies at similar redshift \citep[e.g.,][]{DEugenio2020}.
\par The AGN component in the observed optical/NIR can be estimated from the bolometric luminosity. By using the optical bolometric correction factor $K_{\rm O}\equiv L_{\rm bol}/L_{\rm O}$ of \citet{Duras2020}, here $L_{\rm O}$ is the rest-frame B-band luminosity value, the rest-frame optical luminosity of AGNs is derived. The value is significantly lower than the luminosity of the corresponding observed optical/NIR. For example, the observed average rest-frame B-band luminosity of QGs in the highest redshift bin, derived from the average K band magnitude and the k-correction with the typical SED model of QGs \citep[e.g.,][]{Valentino2020}, is about $\sim30$ times higher than the expected rest-frame optical AGN luminosity. This suggests that the AGN component is much smaller than the stellar component in the optical/NIR, and our SED-fitting assuming only stellar components is well justified.

\subsection{Implication to Galaxy Quenching}\label{sec:5-2}
\par We have systematically investigated the AGN activity of typical QGs based on the X-ray and radio stacking analyses. As shown in Sections \ref{sec:3} and \ref{sec:4}, their X-ray and radio luminosities cannot be explained only by XRBs or SFR, and QGs are found to generally host AGNs with X-ray luminosity of $L_{\rm 2-10keV,AGN}\sim10^{40-43}\ {\rm erg/s}$ at any redshift up to $z\sim5$. The connection between typical QGs and AGNs has been discussed in the literature by focusing on X-ray undetected objects only at $z\sim2$ \citep{Olsen2013}, including the individually detected objects in X-ray at $z<3.5$ \citep[e.g.,][]{Carraro2020,DEugenio2020} and in radio at $z<2$ \citep[e.g.,][] {Magdis2021,Gobat2018,Man2016}. This study only focuses on the X-ray undetected QGs and perform both X-ray and radio analysis consistently out to a previously unreached redshift of $z\sim5$.
\par One of our primary findings is that QGs have higher AGN luminosity than SFGs at $z>1.5$ both in the X-ray and radio. This implies that the quenching tends to occur with more active AGNs, supporting the AGN feedback for quenching massive galaxies at high redshifts. Simulations show the necessity of the AGN feedback to reproduce the observed steep bent of the rest-optical luminosity/stellar mass functions \citep[e.g.,][]{Benson2003,Bower2006,Croton2006,Beckmann2017}. In fact, \citet{Beckmann2017} argue that AGN feedback at higher redshift affects the quenching of even less massive galaxies. According to their best-fit relation, galaxies with $\log{(M_\star/M_\odot)}\geq9.9$ are expected to be largely affected by AGN feedback at $z>1.5$, which is fully consistent with our findings.
\par The AGN feedback is thought to occur mainly through two processes. One is called ``quasar-mode feedback" \citep[e.g.,][] {Silk1998}. In this mode, the wind from AGNs expels gases from galaxies and suppress the star formation. This is thought to occur in high-luminosity AGNs such as QSOs, close to the Eddington limit. The other mode is called ``radio-mode feedback" \citep[also known as kinetic-mode feedback,][]{Fabian1994}. In this mode, low-luminosity AGNs, such as less than one percent of the Eddington luminosity, heats the circumgalactic and halo gas by their radio jets, preventing the gas from cooling. Compared to the quasar-mode feedback, the radio-mode feedback is expected to keep the quiescence rather than reducing the star formation. Considering that AGNs seen in this study are low-luminosity AGNs and the star formation of QGs is already suppressed, we may be witnessing the radio-mode feedback from AGNs in action, which keep the star formation of our QGs suppressed.
\par AGNs are also observed in spectroscopically confirmed massive QGs at $z>1.5$ by other features. Some studies found QGs with broad emission lines \citep{Marsan2015} or report a high [N{\sc ii}]/H$\alpha$ ratio \citep{Kriek2009,Belli2019}, which are likely from AGNs. In addition, there are Spitzer/MIPS detections which cannot be explained by star formation \citep[e.g.,][]{DEugenio2020}. All these results are consistent with our findings. Together with the theoretical support \citep[e.g.,][]{Beckmann2017}, we might probe that AGN activity is a key phenomenon to make massive galaxies in the quiescent phase at $z>1.5$.
\par At $z<1.5$, we do not see such an enhancement of the AGN luminosity of QGs. Their AGN activity is comparable to or even weaker than SFGs. This difference between the high and low redshifts might be due to other quenching mechanisms operating at lower redshifts, such as environment quenching. There is clear observational evidence that the quiescent fraction at $z<1.5$ is higher in denser regions, such as in galaxy clusters or groups, than in the general field \citep[e.g.,][]{Peng2010,Wetzel2012,vanderBurg2013,Kawinwanichakij2017,Nantais2017,Reeves2021}. Several mechanisms, such as ram-pressure stripping, are expected to quench galaxies in high-dense regions more effectively and keep the star formation stopped. At $z>2$, several studies argue that such a trend is not significant \citep[e.g.,][]{Lin2016, Ito2021} or even reversed in terms of SFR \citep[][but also see \citeauthor{Chartab2020} \citeyear{Chartab2020}]{Lemaux2020}, supporting this hypothesis for explaining the different trend. It seems that $z\sim1.5-2$ is the epoch when the significant environmental dependence of galaxy properties emerges in the history of the Universe, and the environment quenching can be the dominant quenching process at lower redshifts. Because they do not necessarily trigger AGN activity, QGs may not exhibit enhanced AGN luminosity. On the other hand, environmental quenching is less dominant at higher redshifts, and instead, the AGN quenching might have the primary role for the quenching, as supported in this study. Thus, the combination of the AGN and the environmental quenching is able to explain our observations.
\par We note that lower redshift bins can tend to have lower X-ray luminosities because we remove X-ray detection from the stacked sample. However, this would not introduce any bias in discussing the redshift evolution of the AGN luminosity difference between QGs and SFGs. We construct the QG and SFG samples in the same way in each redshift bin. Thus, our comparisons between them are fair.
\par To directly conclude its origin, it will be essential to split QGs into multiple subsamples according to other properties, such as their living environment. This requires a larger and uniform galaxy sample. Furthermore, a deeper survey in X-ray and radio is needed. In terms of X-ray observation, the planned Advanced X-ray Imaging Satellite (AXIS) probe and the forthcoming Advanced Telescope for High Energy Astrophysics (ATHENA) will provide an ideal dataset. According to \citet{Marchesi2020} and the average X-ray flux estimated in this study, 0.3Ms AXIS and 14Ms ATHENA observation will detect the soft-band individual signal of QGs even at the highest redshift over $\sim1\deg^2$ and $\sim4\deg^2$, respectively. These new facilities will push us to obtain a clearer connection between the AGN activity and galaxy quenching.
\section{Summary}\label{sec:6}
\par In this paper, we systematically investigate the X-ray and radio properties of typical quiescent galaxies (QGs) with $\log{(M_\star/M_\odot})>10$ at $0<z<5$ for the first time. The QG sample is constructed based on the latest COSMOS2020 catalog with the sSFR criteria. The X-ray and radio stacking analyses are conducted for both QGs and SFGs, and we compare their properties. Our main results are summarized as follows:
\begin{enumerate} 
\item The stacked X-ray flux is detected up to $z\sim5$ for individually non-detected QGs in the most of stellar mass bins with the signal to noise ratio of more than two. The hardness ratios of QGs are tentatively higher than those of SFGs, suggesting higher obscurations.
\item The absorption-corrected rest-frame 2-10keV luminosity of QGs is comparable to that of SFGs for most bins. It increases with increasing stellar mass for both populations.
\item The observed X-ray luminosity is compared with the XRB luminosity expected from the stellar mass and SFR by using the empirical relation of \citet{Lehmer2016}. The observed X-ray luminosity of QGs is significantly higher than the XRB luminosity, especially 5-50 times higher at $z>1$. On the other hand, the X-ray luminosity of SFGs is comparable to or slightly higher ($\leq3$) than the expected XRB luminosity. This trend suggests the existence of the additional radiation source in QGs, i.e., AGN.
\item The X-ray AGN luminosity, defined as the excess of the observed luminosity to the expected XRB luminosity, is estimated. The X-ray AGN luminosity of QGs is higher than that of SFGs at fixed stellar mass at $z>1.5$. This high luminosity implies the possible relationship between the galaxy quenching and the AGN activity. On the other hand, we do not see such enhanced luminosity of QGs at $z<1.5$.
\item The rest-frame 1.4 GHz radio luminosity is also estimated for the same sample. For massive ($\log{(M_\star/M_\odot)}>10.5$) QGs, the stacked radio flux is significantly detected with the signal to noise ratio more than two up to $z\sim5$.
\item Similar to the X-ray analysis, the radio AGN luminosity is derived by subtracting the luminosity related to their star formation estimated from the stellar mass and the SFR by the empirical relation of \citet{Delvecchio2020} from the observed luminosity. Similar to the X-ray luminosity, the only star formation can hardly explain the radio luminosity of QGs, and their radio AGN luminosity is higher than those of SFGs at $z>1.5$. 
\item The X-ray and radio AGN luminosity are well correlated for both QGs and SFGs. The slope of its relation is similar to the local relation in \citet{Merloni2003}. This study is the first time to show that QGs are located at the same scaling relation up to $z\sim5$.
\item  As suggested from the high X-ray AGN luminosity, BHARs of QGs are higher than those of SFGs at $z>1.5$. Moreover, the ${\rm \langle BHAR\rangle/\langle SFR\rangle}$ of QGs is higher than those of SFGs and almost constant as ${\rm \langle BHAR\rangle/\langle SFR\rangle}\sim10^{-3}$ regardless of the redshift.
\end{enumerate}
\par Our study unveils typical properties of AGNs in QGs at $z\sim3-5$ for the first time. The enhanced AGN luminosity at $z>1.5$, revealed independently and consistently from the X-ray and radio, supports a crucial role of AGNs in star formation quenching of massive galaxies, especially through the radio-mode feedback. This is further supported by recent spectroscopic observations of massive QGs, many of which turn out to host AGNs. Less pronounced AGN activity in QGs at $z<1.5$ might be due to the increasing role of environmental quenching at lower redshifts. Our work hints at the evolving role of AGN feedback for galaxy quenching toward higher redshift, and future observations of QGs may shed further light on the detailed physics.
\acknowledgements
We thank the referee, Dr. Suchetana Chatterjee for helpful comments and suggestions in the reviewing process, which helped improve the paper. We acknowledge Dr. Rhythm Shimakawa for helpful discussions. The scientific results reported in this article are partially based on observations made by the Chandra X-ray Observatory. This research has used software provided by the Chandra X-ray Center (CXC) in the PIMMS application package. KI acknowledges support from JSPS grant 20J12461. TM and the development/maintenance of CSTACK are supported by UNAM-DGAPA PAPIIT IN111319 and CONACyT Investigaci\'on Cient\'ifica B\'asica 252531. SM acknowledges funding from the the INAF “Progetti di Ricerca di Rilevante Interesse Nazionale” (PRIN), Bando 2019 (project: “Piercing through the clouds: a multiwavelength study of obscured accretion in nearby supermassive black holes”). The Cosmic Dawn Center (DAWN) is funded by the Danish National Research Foundation under grant No. 140. ST and JW acknowledge support from the European Research Council (ERC) Consolidator Grant funding scheme (project ConTExt, grant No. 648179).
\software{MIZUKI \citep{Tanaka2015},  CSTACK \citep[v.4.32][]{Miyaji2008}, PIMMS \citep{Mukai1993}, Astropy \citep{Robitaille2013,Price-Whelan2018}, Matplotlib \citep{Hunter2007}, numba \citep{Lam2015}, numpy \citep{Harris2020}, pandas \citep{Mckinney2010}}

\appendix
\section{Impact of Removing 24${\rm \mu m}$ Detected Sources}\label{sec:B}
\par
\begin{figure*}
    \centering
    \includegraphics[width=12cm]{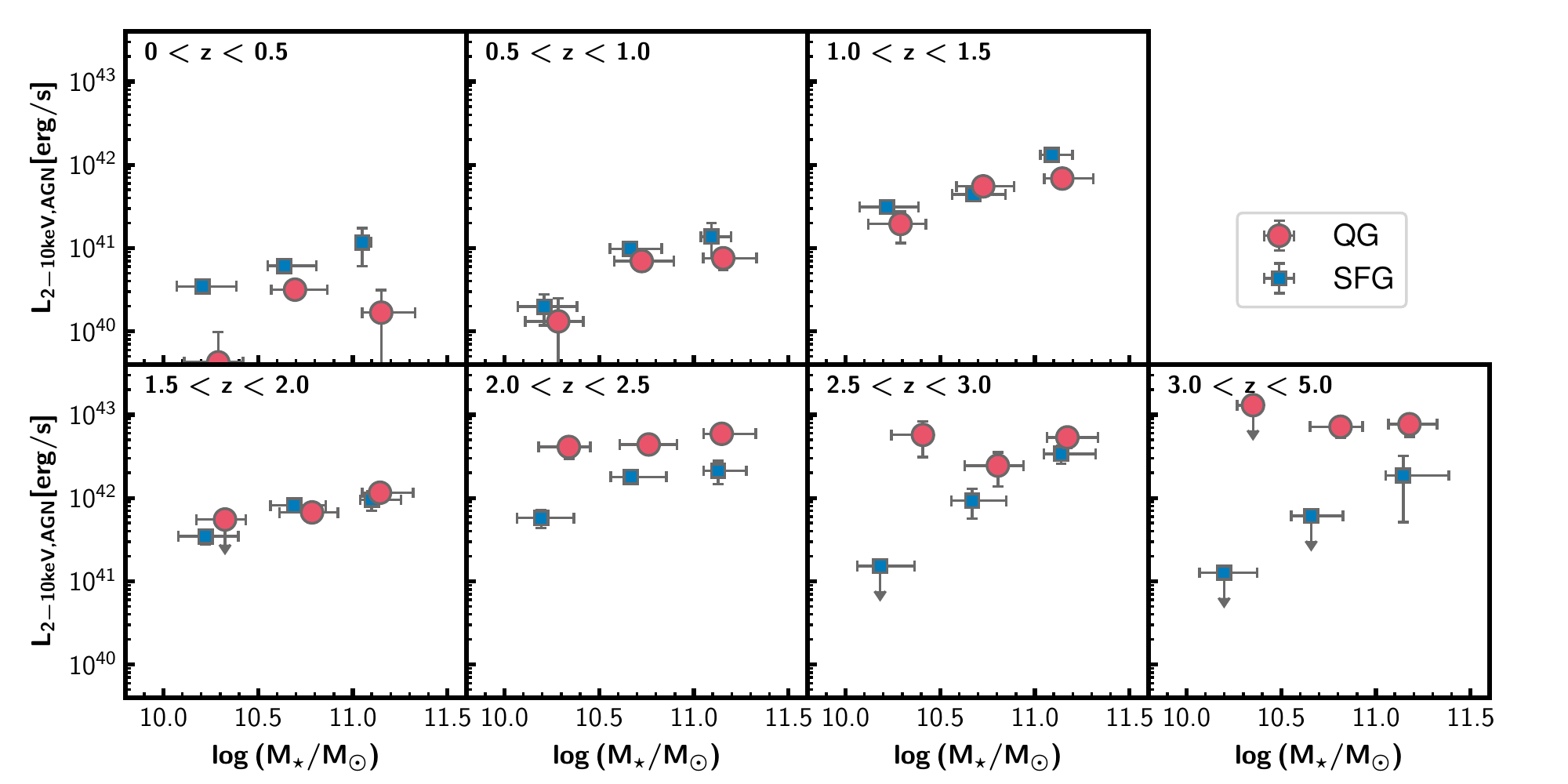}
    \caption{The same for Figure \ref{fig:8}, but in the case of excluding the $24\mu$m detected sources of \citet{LeFloch2009}.}
    \label{fig:17}
\end{figure*}

\begin{figure*}
    \centering
    \includegraphics[width=12cm]{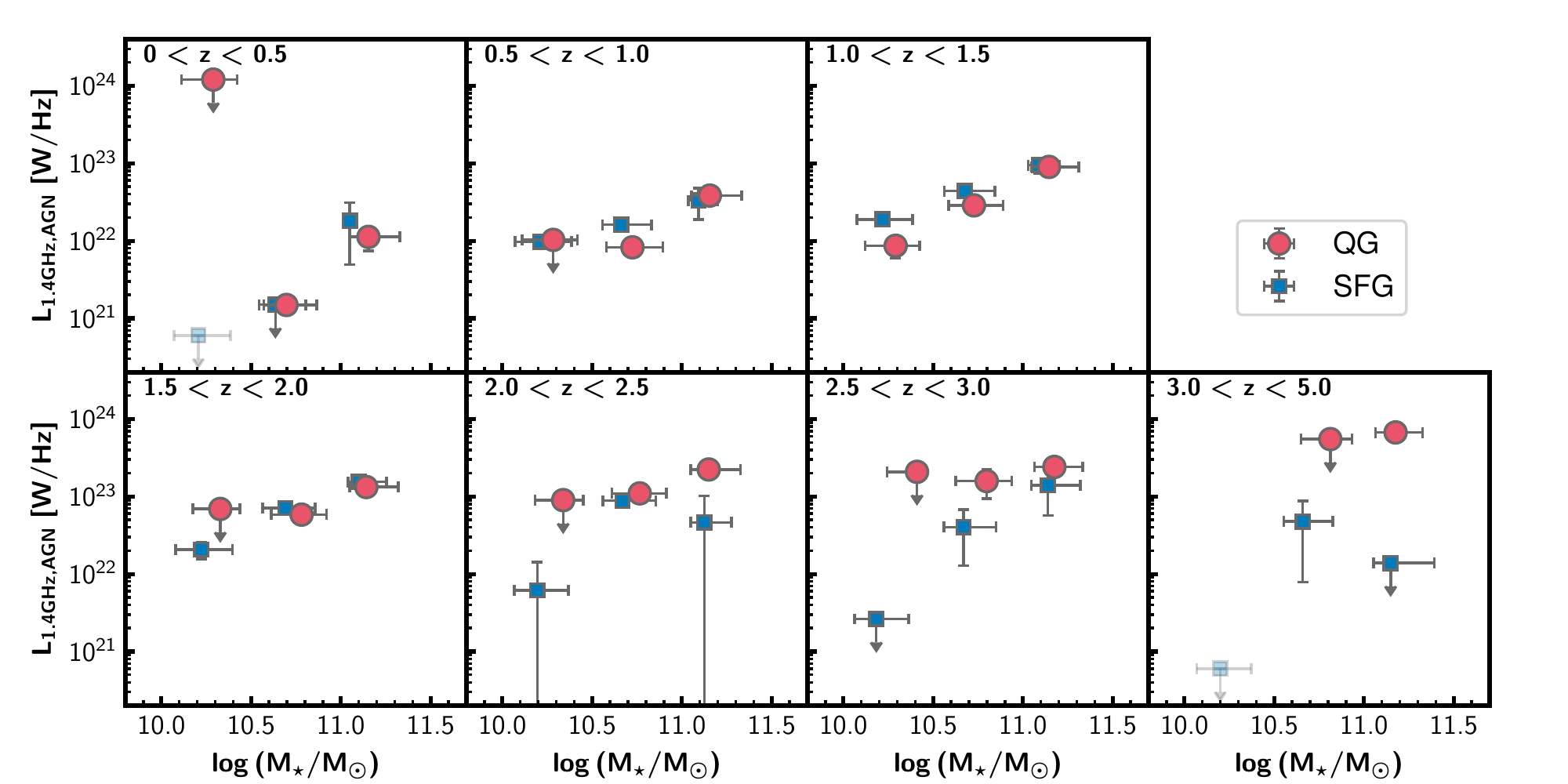}
    \caption{The same for Figure \ref{fig:13}, but in the case of excluding the $24\mu$m detected sources of \citet{LeFloch2009}.}
    \label{fig:18}
\end{figure*}

 Here, we show the result when we remove objects with $S/N>4$ detection in MIPS 24$\mu$m \citep{LeFloch2009} from the sample by allowing the separation of 2\arcsec (see Section \ref{sec:2}). Figure \ref{fig:17} and Figure \ref{fig:18} show the AGN luminosity in the X-ray and radio as a function of redshift and stellar mass, which correspond to Figure \ref{fig:8} and Figure \ref{fig:13} of the original case, respectively. Their trend is the same, i.e., the AGN luminosity of QGs is higher than that of SFGs at $z>1.5$. Therefore, we conclude that the 24$\mu$m detected sources do not affect the entire conclusion of this paper.

\section{Effect of Different XRB Scaling Relation on X-ray AGN luminosity}\label{sec:A}
\par In Section \ref{sec:3}, we estimate the X-ray AGN luminosity from the observed X-ray luminosity by using the XRB scaling relation derived in \citet{Lehmer2016}. As seen in Equation \ref{eq:2}, this relation considers two terms, each proportional to the stellar mass and the SFR. On the other hand, there are other suggested function forms. Especially, \citet{Aird2017} and \citet{Fornasini2018} used the form which set the index of SFR free, which is defined as follows:
\begin{equation}
\begin{split}
        L_{X,\mathrm{XRB}}=\alpha(1+z)^{\gamma} M_{\star}+\beta(1+z)^{\delta} \mathrm{SFR}^ {\theta}.\label{eq:A1}
\end{split}
\end{equation}
\par In, \citet{Aird2017}, the parameters are estimated as $\log{\alpha} = 28.81\pm0.08$, $\log{\beta} = 39.50\pm0.06$, $\gamma = 3.90\pm0.36$, $\delta = 0.67\pm0.31$, and $\theta = 0.86\pm0.05$. Also in \citet{Fornasini2018}, the parameters are estimated as $\log{\alpha} = 29.98\pm0.12$, $\log{\beta} = 39.78\pm0.12$, $\gamma = 0.62\pm0.64$, $\delta < 0.2$, and $\theta = 0.84\pm0.08$. The difference of the method of these two studies is whether the luminosity corrects the intrinsic absorption and the sample selection, such as the redshift range. In addition, it should be noted that \citet{Fornasini2018} mentioned that we should consider the XRB luminosity derived by their best-fit value as the upper limit since their sample can contain low-luminosity AGN. 
\par
\begin{figure*}
    \centering
    \includegraphics[width=12cm]{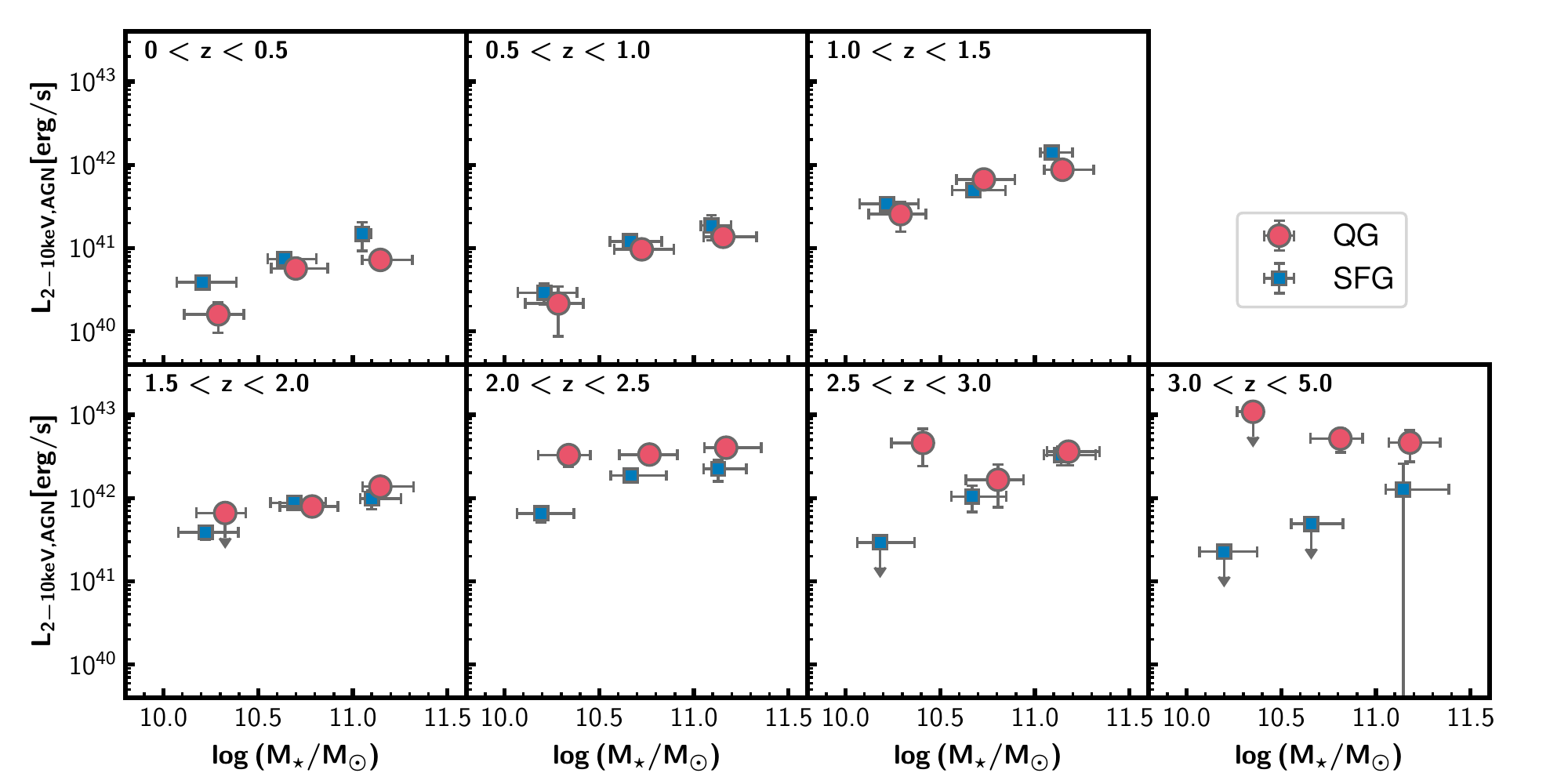}
    \caption{The same for Figure \ref{fig:8}, but in the case of using the XRB scaling relation in \citet{Aird2017}.}
    \label{fig:19}
\end{figure*}

\begin{figure*}
    \centering
    \includegraphics[width=12cm]{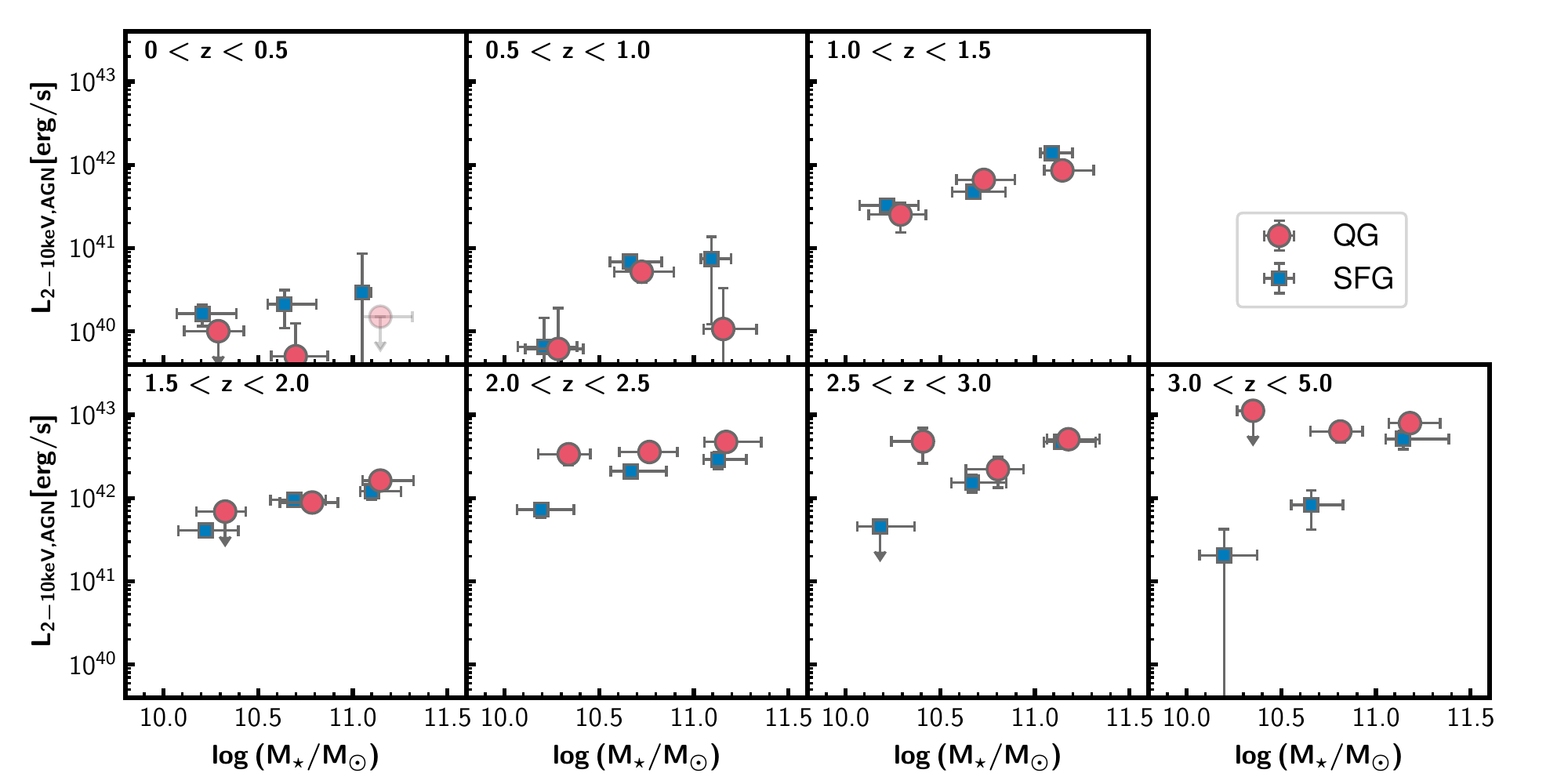}
    \caption{The same for Figure \ref{fig:8}, but in the case of using the XRB scaling relation in \citet{Fornasini2018}. If the upper limit is negative, it is replaced by $1.5 \times 10^{40}$ erg/s only for illustrative purposes (shown in light color).}
    \label{fig:20}
\end{figure*}

 We here use these scaling relations and examine whether the picture obtained in Section \ref{sec:3-4} is changed. The AGN luminosity is derived in the same manner as in Section \ref{sec:3-4}. Figure \ref{fig:19} and Figure \ref{fig:20} show the AGN luminosity in the case of the XRB scaling relation of \citet{Aird2017} and \citet{Fornasini2018}, respectively. Though the exact value changes in both cases, but the overall trend does not change significantly. The AGN luminosity of QGs is generally higher than that of SFGs at $z>1.5$ and their difference gets smaller at $z<1.5$. This result suggests that the trend of the AGN luminosity is robust against the assumed XRB scaling relation. 

\bibliographystyle{aasjournal}

\end{document}